\documentclass[12pt]{article}

\newcommand{\bbr}{I\!\! R}
\newcommand{\bbz}{Z\!\!\! Z}

\newcommand{\3}{$^3$}
\newcommand{\4}{$_4$}

\newcommand{\x}{arXiv:}
\newcommand{\m}{\mathrm}

\usepackage{graphicx}

\begin{document}
\thispagestyle{empty}
\begin{center}

\null \vskip-1truecm \vskip2truecm {\Large{\bf

\textsf{The Most Probable Size of the Universe}

}}

\vskip1truecm {\large \textsf{Brett McInnes}} \vskip1truecm

 \textsf{Abdus Salam ICTP \\ and \\  National
  University of Singapore\footnote{Permanent address}}

email: matmcinn@nus.edu.sg\\

\end{center}
\vskip1truecm \centerline{\textsf{ABSTRACT}} \baselineskip=15pt
\medskip
It has recently been suggested, by Firouzjahi, Sarangi, and Tye,
that string-motivated modifications of the Hartle-Hawking wave
function predict that our Universe came into existence from
``nothing" with a de Sitter-like spacetime geometry and a
spacetime curvature similar to that of ``low-scale" models of
Inflation. This means, however, that the Universe was quite large
at birth. It would be preferable for the initial scale to be close
to the string scale [or perhaps the Planck scale]. The problem
with this, however, is to explain how any initial homogeneity is
preserved during the pre-inflationary era, so that Inflation can
indeed begin. Here we modify a suggestion due to Linde and assume
that the Universe was born with the topology of a torus; however,
we propose that the size of the torus is to be predicted by the
FST wave function. The latter does predict an initial size for the
torus at about the string scale, and it also predicts a
pre-inflationary spacetime geometry such that chaotic mixing
preserves any initial homogeneity until Inflation can begin at a
relatively low scale.

 \vskip3.5truecm
\begin{center}

\end{center}

\newpage

\addtocounter{section}{1}
\section* {\large{\textsf{1. Creating a String Gas Universe From Nothing}}}

One of the most appealing ideas regarding the origin of our world
is that it came into existence out of ``nothing"
\cite{kn:vilenkin1}\cite{kn:vilenkin2}\cite{kn:hartle}\footnote{Clear
reviews of quantum gravity, including many subtleties which will
not be discussed here, are given in
\cite{kn:carlip}\cite{kn:coule}; see also \cite{kn:carlip2} for an
interesting recent development.}. Work on this idea has recently
been resumed, stimulated by developments in string theory
\cite{kn:tye}\cite{kn:laura1}\cite{kn:OVV}\cite{kn:laura2}\cite{kn:dijk}\cite{kn:sarangi}.
In particular, Firouzjahi et al \cite{kn:tye}\cite{kn:sarangi}
take up the question of using the wave function of the Universe to
show that the most probable state of the early Universe is one
that leads to Inflation. Known models of string Inflation [see,
for example, \cite{kn:racetrack}] imply that the inflationary
scale should be several orders of magnitude below the string
scale. In fact, Firouzjahi et al find that their modified wave
function [which takes into account the effects of
\emph{decoherence}] predicts that the Universe should come into
existence, from ``nothing", at about this inflationary scale.

Because ``creation from nothing" tends to be associated with
global de Sitter spacetime, this means in practice that one shows
that the Universe is created in the form of a [perturbed] global
dS\4, with a curvature [proportional to 1/L$^2_{\m{inf}}$] which
corresponds to the inflationary energy scale. Now in dS\4 itself,
L$_{\m{inf}}$ plays two roles: it sets the scale \emph{both} for
the temporal evolution \emph{and} for the size of the spatial
sections [which are spheres, or finite quotients of spheres
\cite{kn:louko1}\cite{kn:rp3}\cite{kn:louko}, of maximal curvature
1/L$^2_{\m{inf}}$]. This corresponds to the fact that
L$_{\m{inf}}$ appears \emph{twice} in the formula for the
[Lorentzian, signature ($+\;-\;-\;-$)] Global de Sitter metric:
\begin{eqnarray}\label{eq:A}
g(\m{GdS_4)(L_{inf})_{+---}\; =\;
d\tau^2\;-\;L_{inf}^2\,cosh^2(\tau/L_{inf})\,[d\chi^2\; +\;
sin^2(\chi)\{d\theta^2 \;+\; sin^2(\theta)\,d\phi^2\}]}.
\end{eqnarray}
In the context of ``creation from nothing" this means that, if we
model the process using de Sitter spacetime, then [topological
complications aside] starting Inflation at a low energy scale
forces the Universe to be born [at $\tau$ = 0] \emph{with a
relatively large size}. It also implies that the shape of the de
Sitter Penrose diagram is fixed: it is square if the spatial
sections are simply connected, rectangular [and precisely twice as
high as wide] if the sections have $\bbr P^3$ topology
\cite{kn:louko1}\cite{kn:rp3}\cite{kn:louko}, and so on.

However, within string theory it seems rather unnatural for the
Universe to be born so large: a size near to the string scale
would seem more appropriate. This is particularly true if we
subscribe to any version of \emph{string gas cosmology}
\cite{kn:brandvafa}, where the string length scale plays a basic
role though T-duality.

Furthermore, as Linde \cite{kn:lindetypical}\cite{kn:lindenew} has
recently emphasised, Inflation necessarily begins with the
Universe in a spatially homogeneous state: \emph{but it is not
easy to understand why the Universe should be created in such a
state if it was born large}. Understanding this question is
fundamental for the approach of Firouzjahi et al, which is based
on the assumption that a cosmological model can only be acceptable
if it passes through an inflationary phase.

These issues are closely related to the deep questions connected
with the apparently extremely ``special" initial conditions of our
Universe, as discussed for example in
\cite{kn:sorbo}\cite{kn:chen}. We certainly will not try to
resolve those problems here, but, as a first step, we note that
they look more tractable if the Universe was born small --- say,
at around the string length scale. For we can then hope that
specifically ``stringy" effects will allow us to explain how the
special initial state was selected. But how can the Universe be
born with a \emph{spatial} size much smaller than its inflationary
\emph{spacetime} scale, L$_{\m{inf}}$?

The solution suggested by Linde [see also \cite{kn:trodden}] makes
use of the proposal of Zel'dovich and Starobinsky \cite{kn:zelda},
who considered the possibility that the Universe began [perhaps as
a Planck-scale fluctuation] with \emph{compact, flat} spatial
sections instead of locally spherical ones --- that is, the
spatial sections have the topology of a three-dimensional torus or
of one of the non-singular quotients of a three-torus
\cite{kn:ruback}\cite{kn:reallyflat}. We can explain the relevance
of this idea in the following way: in such cosmologies, \emph{the
spatial scale is decoupled from the spacetime curvature scale.}
That is, if the spatial tori\footnote{For simplicity, we take the
spatial sections to be cubic tori. Everything we shall say applies
equally to tori of other shapes and to quotients of tori; the
reader merely has to interpret ``length scale" in the appropriate
way.} are characterized by a length K, this K can determine the
initial size without having any necessary relationship with the
spacetime curvature scale L. For an explicit example, take the
version of de Sitter spacetime with a spacetime curvature length
scale L and a spatial section at time t = 0 which is a cubic torus
of side length 2$\pi$K. The metric of this Spatially Toral de
Sitter spacetime is
\begin{equation}\label{eq:B}
g(\m{STdS})(\m{K,\,L)_{+---} \;=\; dt^2\;
-\;K^2\,e^{(2\,t/L)}\,[\,d\theta_1^2 \;+\; d\theta_2^2 \;+\;
d\theta_3^2]},
\end{equation}
where the torus is parametrized by angles which run from $-\,\pi$
to $+\,\pi$. [This is the metric used by Ooguri et al
\cite{kn:OVV} in their recent study of the duality between black
holes and accelerating cosmologies.] Clearly K and L are
\emph{entirely independent}, and this will continue to be the case
for metrics of this kind on $\bbr\;\times$ T\3 with more general
scale functions.

Linde's suggestion is that the Universe began, at the beginning of
a \emph{pre-inflationary phase}, with flat, compact spatial
sections of length scale K, and with a spacetime curvature scale
L$_{\m{inf}}$. However, the fact that K is independent of L opens
up the possibility that K is significantly smaller than
L$_{\m{inf}}$. We wish to consider the possibility that K should
approximate the string scale. Linde argues, by contrast, that the
pre-inflationary era begins at about the \emph{Planck} scale.
There are certain advantages to that proposal: Linde stresses that
in this way we can take full advantage of the decoupling of
spatial and spacetime scales --- with flat, compact sections there
is no need for any tunnelling at all, so that the creation of such
a spacetime is not suppressed. While this proposal is as simple as
possible, we believe that starting at the string scale is also
worth investigating; for, as mentioned earlier, it is natural from
the point of view of string gas cosmology \cite{kn:brandvafa}, in
which the Universe is effectively never smaller than roughly the
string scale. The great virtue of this approach is that it allows
us to give a stringy resolution of the initial singularity. As we
shall review later, singularities are otherwise very hard to avoid
in toral cosmologies, so this is a considerable advantage.

In either case, whether the initial scale is Planckian or stringy,
the Universe expands from the initial state until it reaches a
size comparable to L$_{\m{inf}}$, at which point Inflation is
supposed to begin.

The problem with this idea is that, even if we were to succeed in
solving the deep problem of ``specialness" at the string [or
Planck] scale, we need somehow to \emph{maintain} homogeneity
until the Universe has expanded to the inflationary scale, since
otherwise Inflation will not be able to begin. However, Linde's
proposal can automatically take care of this problem,
\emph{provided that the pre-inflationary spacetime geometry has a
suitable form}. For if the geometry is such that all parts of the
Universe remain in causal contact during the pre-inflationary
phase, then any initial homogeneity can then be preserved by
causal processes [``chaotic mixing" \cite{kn:mixing}] until this
\emph{global causal contact} is lost.

We thus obtain a very appealing picture: the Universe is born,
either from a Planckian fluctuation or from ``nothing", in a very
homogeneous state, and with nearly [or even \emph{exactly}] flat,
compact spatial sections. The spacetime curvature scale is already
L$_{\m{inf}}$, but the spatial scale has a \emph{significantly
smaller} value, K. The Universe then expands until the
inflationary scale is reached, at which point Inflation will begin
if all spatial gradients are [still] sufficiently small by then.
But this last condition can be met if the pre-inflationary
spacetime geometry allows global causal contact until that time.
We refer to this picture as ``Linde's programme", though we remind
the reader that Linde argues for a value of K different from the
one to be discussed here.

However, the basic questions now are these: what fixes the value
of K relative to L$_{\m{inf}}$? How is the correct
pre-inflationary spacetime geometry chosen? Only if we can answer
these questions can we claim to have implemented Linde's programme
for low-scale Inflation.

If we insist on a certain mild energy condition, to be discussed
below, then a FRW cosmology with flat spatial sections cannot be
``created from nothing", and we are forced back to the Planck
length for K, as Linde advocates. The drawback here is that the
origin of the Universe, and the way in which the initial
singularity is resolved, can only be described in terms of
Planck-scale physics of which we still have little or no
understanding.

Here we shall explore what seems to be the only alternative:
\emph{we apply the FST wave function to the case where the
Universe tunnels from ``nothing" to an initial state with compact
but flat, instead of locally spherical, spatial sections}. We
assume that the inflationary scale is fixed by string theory
\cite{kn:racetrack} and focus on determining the most probable
value of K, and the most probable spacetime geometry in the
earliest Universe. The hope, of course, is to realize the scenario
described above.

It should be clear that this is an ambitious undertaking. It is
not obvious that merely changing the topology of the spatial
sections will have the desired strong effect on the initial scale,
changing it from the inflationary scale to the string scale.
Likewise, it is far from obvious that, even if we achieve this,
the pre-inflationary spacetime geometry will be such as to
maintain chaotic mixing long enough to allow Inflation to start.
We shall find, however, that the FST wave function \emph{does
predict precisely these things}.

Although it is clear that de Sitter spacetime itself will not
allow us to implement Linde's programme, it is still possible that
we might be able to use a sufficiently deformed version of dS$_4$
to do so. We begin with a discussion of this possibility and its
difficulties; this discussion allows us to clarify, in terms of
the Penrose diagram, the detailed technical requirements which we
shall have to satisfy in order for Linde's programme to work. We
conclude that it is simpler to try toral sections. We then briefly
summarize the singularity theory of cosmological models with toral
spatial sections, emphasising that they can only be created from
``nothing" if a certain geometric condition, the \emph{Null Ricci
Condition}, is violated. In the next section we use this fact to
guide us in constructing explicit examples of simple cosmological
models with toral sections which can be created from ``nothing".
These form a single-parameter family. We then compute the FST wave
function for this family and use it to predict the most probable
values of both the family parameter --- so that we know which
geometry we are [most probably] dealing with
--- and, with reasonable assumptions regarding the string length
scale, of K/L$_{\m{inf}}$. Though the model is of course
over-simplified, the result is quite satisfactory, in the sense
that K/L$_{\m{inf}}$ is predicted to be small --- putting K at
about the string scale --- and the predicted spacetime geometry
does lead to global causal contact which is lost only at about the
time Inflation begins. Furthermore, there is reason to hope that
more realistic versions will continue to yield satisfactory
results.

\addtocounter{section}{1}
\section* {\large{\textsf{2. Locally Spherical Spatial Sections}}}
Before turning to the case of flat, compact spatial sections, let
us first discuss the possibility of using locally spherical
spatial sections to realize Linde's scenario. Note that locally
spherical spatial sections are the standard choice in discussions
of creation from ``nothing" \cite{kn:tye}\cite{kn:sarangi}, and,
as we shall see, there is a sense in which this is the most
conservative choice; so we should explain why we shall \emph{not}
proceed in this way in this work. Doing so will allow us to
explain the technical requirements for Linde's proposal in this
more familiar context.

Simply connected de Sitter spacetime with [in the signature we use
here] spacetime curvature $-$1/L$^2$ is defined as the locus, in
five-dimensional Minkowski spacetime [signature
($+\;-\;-\;-\;-$)], defined by the equation
\begin{equation}\label{eq:C}
\m{+\; A^2\; - \;W^2 \;- \;Z^2\; -\; Y^2\; - \;X^2\; =\; -\;L^2}.
\end{equation}
This locus has topology $\bbr\times\,\m{S}^3$, and it can
therefore be parametrized by \emph{global} conformal coordinates
($\eta,\,\chi,\,\theta,\,\phi)$ defined by
\begin{eqnarray} \label{eq:D}
\m{A} & = & \m{L\;cot(\eta)  }                     \nonumber \\
\m{W} & = & \m{L\;cosec(\eta)\;cos(\chi)}                     \nonumber \\
\m{Z} & = & \m{L\;cosec(\eta)\;sin(\chi)\;cos(\theta)}           \nonumber \\
\m{Y} & = & \m{L\;cosec(\eta)\;sin(\chi)\;sin(\theta)\;sin(\phi)}  \nonumber \\
\m{X} & = & \m{L\;cosec(\eta)\;sin(\chi)\;sin(\theta)\;cos(\phi)}.
\end{eqnarray}
Here $\chi,\,\theta,\,\phi$ are the usual coordinates on the
three-sphere, and $\eta$ is angular conformal time, which takes
its values in the interval ($0,\;\pi$). [Conformal time itself is
then given by $\eta\,$L.] The metric of Global de Sitter spacetime
is then
\begin{equation}\label{eq:E}
g(\m{GdS_4)(L)_{+---}\; =\; L^2\,cosec^2(\eta)[ \; d\eta^2 \; -\;
d\chi^2 \;-\; sin^2(\chi)\{d\theta^2  \;+\;
sin^2(\theta)d\phi^2\}}].
\end{equation}
An obvious conformal transformation allows us to extend the range
of $\eta$, so that it takes all values in [$0,\;\pi$]. The Penrose
diagram is clearly square in the case of \emph{simply connected}
spatial sections, since $\chi$ also has this range in that case.
[It will be narrower in the non-simply-connected case, since, for
example, $\chi$ only runs to $\pi$/2 on $\bbr P^3$.]

Let us briefly recall how to use this diagram to explain how
Inflation solves the horizon problem. In Figure 1, the triangular
\begin{figure}[!h]
\centering
\includegraphics[width=0.7\textwidth]{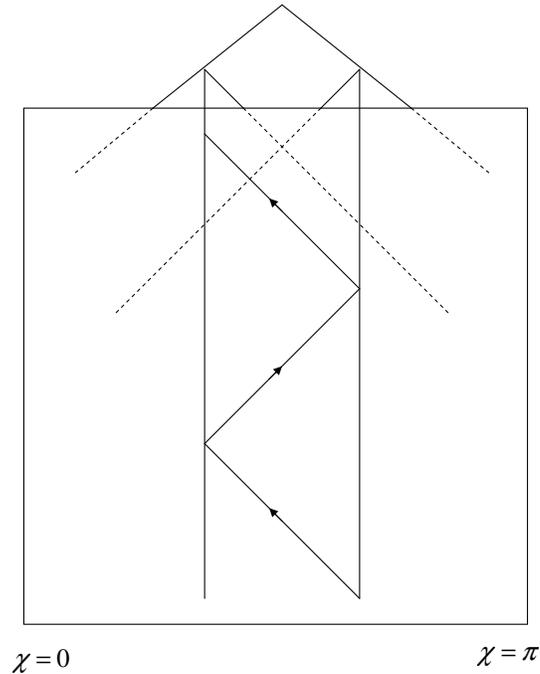}
\caption{How Inflation Solves the Horizon Problem.}
\end{figure}
structures above the upper horizontal line [which schematically
represents both the end of Inflation and decoupling] give the
usual representation of the horizon problem: events observed in
opposite directions apparently have no common past. Below the
line, deeper into de Sitter spacetime, however, we see that the
two events do in fact share much of their past. Furthermore,
inside the shared region there has been ample opportunity for
objects with worldlines passing through these events to exchange
signals. This explains the similarity of the conditions observed
at the two events in question. Notice that the explanation has a
definite ``holographic" flavour: the behaviour on the de Sitter
boundary is understood in terms of interactions in the bulk.
Notice too that the topology of the spatial sections is irrelevant
to this argument: it would not matter, for example, if the Penrose
diagram were rectangular rather than square, as would be the case
if the spatial sections were assumed to be copies of $\bbr P^3$.
[Of course, the size of the post-inflationary part of the diagram
has been enormously exaggerated in Figure 1.]

This explanation works because the events we see in opposite
directions are, by inflationary standards, not very remote from
each other. The explanation would be considerably less convincing
if we were discussing objects at $\chi$ = 0 and $\chi$ = $\pi$.
For, even asymptotically, these objects share only half of their
past events in the simply connected case, and there is \emph{no}
opportunity for signals to be exchanged even in that region.
Actually, the situation here is even worse than it appears. For
global de Sitter spacetime has both a contracting phase and an
expanding phase. We usually delete the former, on the grounds that
it will be cut off either by a singularity or simply in the course
of constructing the usual Hartle-Hawking Euclidean-to-Lorentzian
transition. In this case we should retain only the upper half of
the Penrose diagram. In the simply connected case, this would mean
that antipodal observers have \emph{no} past events in common, and
they are completely out of causal contact. Clearly, simply
connected de Sitter spacetime would not allow us to explain
homogeneity on a full spacelike slice, if such an explanation were
required.

The point of this discussion is that if we are hoping to use a
pre-inflationary era to prepare suitable conditions for the
beginning of Inflation, as Linde proposes
\cite{kn:lindetypical}\cite{kn:lindenew}, and if we insist on
using a spacetime with locally spherical spatial sections, then
the spacetime geometry will have to be very different from that of
de Sitter spacetime.

During the pre-inflationary era, however, the inflaton energy does
not yet completely eclipse other forms of energy. We have not yet
taken into account the effect of these other forms of energy on
the spacetime geometry. Now a theorem of Gao and Wald
\cite{kn:gaowald} implies that if a spacetime is globally
hyperbolic [with compact sections] and null geodesically complete,
and if it satisfies the Null Energy Condition [NEC] and the null
generic condition\footnote{This is the requirement that, along
every null geodesic, there should exist a point where the tangent
vector k$^{\m{a}}$ and the curvature R$_{\m{abcd}}$ satisfy
$\m{k_{[a}\,R_{b]cd[e}\,k_{f]}\,k^c\,k^d \neq 0}$. Exact de Sitter
spacetime itself does \emph{not} satisfy this condition, which is
why the Gao-Wald conclusion is not true of it.}, then any observer
will be able to ``see" an entire spacelike slice at some
sufficiently late time --- which is \emph{not} possible in exact,
simply connected de Sitter spacetime. Essentially what happens in
such a spacetime is that null geodesics sent out from any point
are able to ``turn around" and begin to return towards that point.
[One says that the spacetime has \emph{no null lines}
\cite{kn:gregnull}: on every inextensible null geodesic there
exists a pair of points with timelike separation.] This means that
the introduction of any kind of generic matter satisfying the NEC
will cause the simply connected de Sitter Penrose diagram to
become taller than it is wide, so that null geodesics can
``bounce" off one or both of the vertical edges of the diagram, as
discussed in \cite{kn:smash}\cite{kn:leblond}\cite{kn:unstable}.

The Gao-Wald theorem only describes the \emph{ratio} of the height
to the width of the perturbed de Sitter Penrose diagram; it does
not tell us how the actual minimal size of a spatial section is
affected. In fact, however, the effect of introducing matter which
satisfies the NEC is to make the Universe \emph{smaller} at its
minimum size: that is, if the spatial sections are spheres, the
minimum possible radius is L in de Sitter space, but it is less
than L in the perturbed spacetime. This is related to the tendency
of such matter to focus null geodesics. In the case in which the
matter is introduced in such a way as to preserve the FRW
character of the spacetime, this can be seen directly from the
Friedmann equation, which, with Distorted de Sitter metric
\begin{eqnarray}\label{eq:CEYLON}
g(\m{DdS)(L)_{+---}\; =\; d\tau^2\;-\;L^2\,a^2(\tau/L)\,[d\chi^2\;
+\; sin^2(\chi)\{d\theta^2 \;+\; sin^2(\theta)\,d\phi^2\}]}
\end{eqnarray}
takes the form
\begin{equation}\label{eq:COLOBUS}
\m{L^2\,\dot{a}^2\;=\;{{8\pi
L_P^2}\over{3}}\,\rho\,L^2\,a^2\;-\;1,}
\end{equation}
where a($\tau$) is the dimensionless scale function\footnote{Note
that, with these conventions, the minimal value of the scale
function is \emph{not} equal to unity, except in the case of pure
de Sitter spacetime.}, L$_{\m{P}}$ is the Planck length, L is the
length scale in the initial de Sitter spacetime, and $\rho$ is the
total energy density [de Sitter energy density plus the matter we
are introducing into de Sitter spacetime]. Our objective is to
deform de Sitter spacetime, but not to the extent of causing it to
become singular
--- that would rule out creation from ``nothing". Therefore we
assume that a($\tau$) still has a minimum value, $\m{a_{min}}$.
From (\ref{eq:COLOBUS}) we have
\begin{equation}\label{eq:CROC}
\m{{{8\pi L_P^2}\over{3}}\,\rho(a_{min})\,L^2\,a_{min}^2 \;-
\;1\;=\; 0}.
\end{equation}
Writing $\m{\rho\;=\;3/(8\pi L_P^2 L^2)\;+\; \rho_*}$, where
$\m{3/(8\pi L_P^2 L^2)}$ is the energy density of de Sitter
spacetime and where $\rho_*$ is the energy density we are
introducing, we have now
\begin{equation}\label{eq:CAT}
\m{1\;-\; a_{min}^2 \;=\; {{8\pi
L_P^2}\over{3}}\,\rho_*(a_{min})\,L^2\,a_{min}^2};
\end{equation}
clearly the minimum value of the scale factor is smaller than its
value in de Sitter spacetime, which of course is unity. Thus the
minimum radius, $\m{a_{min}}$L, is indeed smaller than the de
Sitter value, L.

The metric now is still conformal to the standard metric [see the
bracketed expression on the right side of equation (\ref{eq:E})]
on the product of a closed interval with a local three-sphere, so
the Penrose diagram remains rectangular and of the same width;
but, as we explained above, the Gao-Wald theorem implies that the
height of the diagram must increase. Hence the total angular
conformal time [starting from the minimal radius section, since we
are not interested in a contracting phase] always exceeds
$\pi\,$/2.

Thus we see that the effect of taking into account the
back-reaction of the matter content is to make the de Sitter
minimal radius smaller and to make the Penrose diagram taller.
Since the density of non-inflaton matter in the pre-inflationary
phase is relatively large [compared to the inflationary era, when
it is completely dominated by the inflaton energy], this effect
can be very important.

\begin{figure}[!h]
\centering
\includegraphics[width=0.7\textwidth]{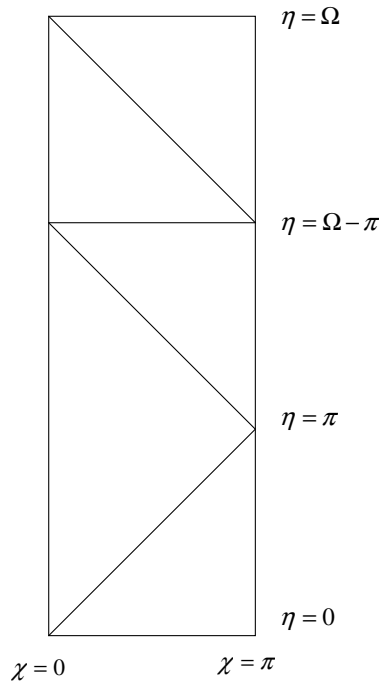}
\caption{Penrose diagram for the pre-inflationary/inflationary
Eras.}
\end{figure}

\emph{It should be clear from our discussion above that small
initial radii and tall Penrose diagrams are precisely what we
require in order to carry out Linde's programme for low-scale
Inflation}. For of course the ``tallness" will allow particles and
signals to be sent back and forth between all parts of the
Universe. The era during which this is possible is what we call
the ``pre-inflationary" era. Causal contact begins to be lost when
the cosmological horizon begins to develop. Since Inflation can
only begin if the matter distribution is sufficiently homogeneous,
\emph{it is essential that the Universe should have reached the
Inflationary scale by this time.} We shall refer to the subsequent
phase as the inflationary era. This is the top square in Figure 2;
that is, the inflationary era must begin roughly $\pi$ units of
angular conformal time below the upper boundary of the Penrose
diagram [which is at angular conformal time $\eta\;=\;\Omega$].

Thus Figure 2 portrays Linde's programme: the pre-inflationary era
must be such that homogeneity is maintained until the surface
$\eta\;=\;\Omega\,-\,\pi$ is reached, which is roughly when
Inflation itself should begin; the reader can picture this line as
a compact hypersurface of typical ``inflationary size". As a
general summary of Linde's programme, this diagram [with suitable
changes of coordinate labels] is valid for any spatial topology,
provided that the sections are \emph{compact}. In particular, this
same diagram gives the conditions for the programme to work in the
case of \emph{toral} spatial sections\footnote{$\chi$ will in that
case be replaced by one of the angular coordinates on the torus,
taken to run from $-\pi$ to $+\pi$.}.

The question is now: how small can the initial radius be, and how
tall can the Penrose diagram become, in the case of locally
spherical spatial sections?

Physically, one might reason as follows. Clearly, there must be a
bound, imposed by self-consistency, on how much [conventional]
matter can be born with a Universe which is created from
``nothing". For if there is too much matter initially, it will
dominate the inflaton to such an extent that the Strong Energy
Condition [SEC] will be satisfied \emph{initially}, even if it is
violated later when the conventional matter dilutes. The classical
singularity theorems \cite{kn:hawking}, applied to this early
phase alone, will require the spacetime to be singular --- which
of course would mean that the Universe was \emph{not} born from
``nothing". Given that the introduction of conventional matter
causes the minimum radius to shrink and the total conformal time
to increase, one would expect to be able to arrange to have an
arbitrarily small initial radius, and an arbitrarily tall Penrose
diagram, by carefully adjusting the amount of matter to be
sufficiently close to the critical value. \emph{For locally
spherical sections, only the second part of this statement is
correct, however.}

To see this, let us consider a concrete example. To be specific,
we shall assume that the Universe is born with locally spherical
structure, with a positive cosmological constant having energy
density $\rho_{\Lambda}$, and containing some kind of radiation
with initial density $\rho_{\m{0R}}$. As always in discussions of
creation from ``nothing", we assume that the
Euclidean-to-Lorentzian transition takes place along a surface of
zero extrinsic curvature, that is, a surface of minimal size: so
$\rho_{\m{0R}}$ is also the \emph{maximal} radiation density. The
assumption that the matter takes the form of radiation is
motivated by \cite{kn:sarangi}, where indeed it is found that the
modified wave function of the Universe naturally leads to the
creation of exactly this kind of spacetime. We shall find it
useful to define a parameter
\begin{equation}\label{eq:CAMERA}
\kappa \;=\;\rho_{\m{0R}}/\rho_{\Lambda}.
\end{equation}
Now because radiation satisfies the SEC, there is an upper bound
on $\kappa$; otherwise, as explained above, the combination of the
cosmological constant and the radiation will also satisfy the SEC
initially, and then the classical singularity theorems [applied to
the earliest phase, before the radiation energy density drops to
the point where the SEC is violated] will imply the presence of an
initial singularity. The SEC is violated initially [and therefore
at all times] provided that
\begin{equation}\label{eq:CUCKOO}
[\rho_{\m{0R}}\;+\;
\rho_{\Lambda}]\;+\;3[{{1}\over{3}}\,\rho_{\m{0R}}\;-\;\rho_{\Lambda}]\;<\;0,
\end{equation}
where we have used the respective equations of state for the
cosmological constant and for radiation\footnote{Actually, the
spacetime remains non-singular even if this inequality is
saturated. But then the extrinsic curvature of a spatial slice is
never zero --- it only approaches zero asymptotically to the
past.}. This simply means that we must have
\begin{equation}\label{eq:CROSSFIRE}
\kappa \;<\;1
\end{equation}
if we are to deform de Sitter spacetime without causing a
singularity.

Assuming now that we are dealing with a FRW model described by
equation (\ref{eq:COLOBUS}), we can write the radiation density as
\begin{equation}\label{eq:CURRY}
\m{\rho_R(\tau)\;=\;{{3\,\alpha}\over{8\pi
\,L_P^2\,L^2}}\,a^{-\,4}},
\end{equation}
where $\alpha$ is a positive constant. Since the energy density
associated with the cosmological constant is
3/8$\pi$L$_{\m{P}}^2$L$^2$, we have, from equation
(\ref{eq:COLOBUS}),
\begin{equation}\label{eq:CHEESE}
\m{L^2\,\dot{a}^2\;=\;\alpha \,a^{-2}\;+\;a^2\;-\;1}.
\end{equation}
Thinking of the right side as an effective potential, and noting
again that we are interested in spacetimes which are perturbations
of de Sitter spacetime [in the sense that, like the latter, they
have a minimum value of the scale function rather than a maximum],
one sees that the minimum value of a($\tau$) is found by taking
the larger positive root of a$^4$ $-$ a$^2$ + $\alpha$; that is,
\begin{equation}\label{eq:CRACKERS}
\m{a_{min}\;=\;\sqrt{{{1}\over{2}}[1\;+\;\sqrt{1\;-\;4\alpha}\,]}}.
\end{equation}
Notice that there is an upper bound on $\alpha$; this will turn
out to be equivalent to (\ref{eq:CROSSFIRE}), that is, it just
expresses the requirement that the SEC should be violated at all
times.

Since the maximum density occurs when a($\tau$) is smallest,
substituting (\ref{eq:CRACKERS}) into (\ref{eq:CURRY}) allows us
to express the maximal radiation density, which we denoted
$\rho_{\m{0R}}$ above, in terms of $\alpha$. Solving for $\alpha$,
we can express it in terms of $\rho_{\m{0R}}$ and thus in terms of
$\kappa$:
\begin{equation}\label{eq:CHEDDAR}
\m{\alpha\;=\;{{\kappa}\over{[1\;+\;\kappa\,]^2}}}.
\end{equation}
Substituting this into (\ref{eq:CRACKERS}) we have an expression
for the minimum value of the scale function in terms of $\kappa$:
\begin{equation}\label{eq:CRAP}
\m{a_{min}\;=\;{{1}\over{\sqrt{\,1\;+\;\kappa}}}}.
\end{equation}
Notice that, as expected, the minimum value of a($\tau$) is always
less than its de Sitter counterpart, unity.

Provided that (\ref{eq:CROSSFIRE}) is satisfied, the Penrose
diagram in the simply connected case will be a rectangle of width
$\pi$; the height will be given by the full extent of [angular]
conformal time, beginning at the start of the Lorentzian era; we
denote this height by $\Omega_{\kappa}$. This is easily computed
by solving (\ref{eq:COLOBUS}) for d$\eta$ = dt/La($\tau$) and
integrating:
\begin{equation}\label{eq:COCONUT}
\m{\Omega_{\kappa}\;=\;\int_{a_{min}}^{\infty}{{da}\over{\sqrt{a^4\;-\;a^2\;+\;\alpha}}}},
\end{equation}
where $\alpha$ is given by (\ref{eq:CHEDDAR}) and a$_{\m{min}}$ is
given by (\ref{eq:CRAP}). It is important to bear in mind that
indeed a$_{\m{min}}$ does depend on $\kappa$; otherwise
(\ref{eq:COCONUT}) would give the impression that the introduction
of matter [positive $\alpha$] \emph{shortens} the Penrose diagram,
which, as we know, is never the case for matter with a positive
density. Notice finally that $\Omega_{\kappa}$ is the angular
conformal time expended during the \emph{expansion} of this
spacetime: the contracting phase is cut away, since it is to be
replaced by a Euclidean space in the construction of the wave
function. Thus in particular $\Omega_0$ = $\pi$/2, as in the
\emph{expanding} phase of simply connected de Sitter, \emph{not}
$\pi$.

We are now in a position to answer our questions. First, we see at
once from (\ref{eq:CROSSFIRE}) and (\ref{eq:CRAP}) that,
unfortunately, it is \emph{not} possible to reduce the initial
size of the Universe very substantially by assuming that it is
born containing radiation. Even in the limiting case, in which
$\kappa$ approaches unity, the reduction of the radius is only to
approximately 71\% of its de Sitter value. If the latter has its
usual inflationary value, then the Universe is \emph{not} born at
the string scale [which is usually estimated to be about two
orders of magnitude smaller than this], no matter how much
radiation is born with it.

On the other hand, if we let $\kappa$ $\rightarrow$ 1 in
(\ref{eq:COCONUT}), the effect will be to cause the two roots of
the quadratic in a$^2$ to coincide, causing a logarithmic
divergence; however, the integral always converges for any value
of $\kappa$ strictly less than unity. Thus, $\Omega_{\kappa}$ can
be made \emph{arbitrarily large} by choosing $\kappa$ sufficiently
close to unity. Hence we can make the Penrose diagram as tall as
we wish. Unfortunately, however, large values of $\Omega_{\kappa}$
can be achieved only at the cost of a fairly severe fine-tuning,
because in fact $\Omega_{\kappa}$ grows extremely slowly as
$\kappa$ approaches unity. Thus we have to think carefully about
the extent of vertical stretching we need to achieve in order to
achieve chaotic mixing, as in Linde's programme.

Recalling the way we used the Penrose diagram [Figure 1] to
explain the workings of conventional Inflation, we must at least
require that the pre-inflationary era should belong to the shared
past of antipodal observers, and that it should be possible for
these observers to exchange at least one signal during that era.
[In reality, of course, we would want more communication than
this, but we are seeking here to establish a lower bound.] For
that, we must stretch the ``expanding" [upper] half of the simply
connected de Sitter Penrose diagram by a factor of at least 6
--- see Figure 2
--- so that its height becomes 3$\pi$. This allows for precisely
one complete circumnavigation during the pre-inflationary era. But
a numerical investigation of the integral in (\ref{eq:COCONUT})
shows that, even to achieve this most conservative case, one needs
$\kappa$ to be at least 0.99997. Combining this with
(\ref{eq:CROSSFIRE}), we have
\begin{equation}\label{eq:CRABS}
0.99997 \;<\; \kappa \;<\;1.
\end{equation}
If one wishes to ensure that it should be possible for signals and
objects to perform multiple circumnavigations in this era, then of
course the fine-tuning becomes substantially worse.

In short, it is not easy to realise Linde's pre-inflationary
scenario using locally spherical spatial sections. This is
\emph{not} to say that it cannot be done. For example, one could
try to obtain smaller initial radii by considering Universes born
with other kinds of matter apart from radiation. This can in fact
be made to work. Furthermore, the Penrose diagram can be made
substantially narrower by considering topologically non-trivial
versions of de Sitter spacetime, such as the one, mentioned
earlier, with $\bbr P^3$ spatial sections. There is of course no
physical justification whatever for assuming that global de Sitter
spacetime should be simply connected; on the contrary, there are
very good reasons for supposing that it isn't
\cite{kn:rp3}\cite{kn:orbifold}. For global de Sitter spacetime
[with locally spherical sections] this means that the topology of
the underlying manifold is taken to be $\bbr
\;\times\;[\m{S}^3/\Gamma$], where $\Gamma$ is a finite group from
a well-known list \cite{kn:weeksphere}.

Apart from $\bbr P^3$, the manifolds $\m{S}^3/\Gamma$ are not
globally isotropic, and most of them are not homogeneous, so,
strictly speaking, one cannot draw a Penrose diagram for the
corresponding versions of global de Sitter spacetime; nevertheless
we can argue as follows. First, in view of our objectives here, it
is reasonable to confine attention to the homogeneous quotients of
S$^3$; furthermore, we are not interested in spaces where the
quotient is smaller than S$^3$ in some directions but not in
others [as is the case for most of the lens spaces]. The obvious
candidate is the well-known binary icosahedral quotient
S$^3$/I$^*_{120}$, where I$^*_{120}$ is the group, of order 120,
which is the double cover of the group of symmetries of a regular
icosahedron. For a given curvature, S$^3$/I$^*_{120}$ is up to ten
times smaller than S$^3$; that is, it is ten times smaller in
certain directions, but the reduction is less dramatic in other
directions. As the reduction in volume is clearly by a factor of
120, we can use five as an estimate of the reduction in the width
of the ``Penrose diagram" for this version of de Sitter spacetime.
Thus the diagram is somewhat taller than wide, even before we take
into account the presence of matter.

Combining all these observations, we can put together a picture of
the kind proposed by Linde, while using locally spherical spatial
sections. That is, we can construct a pre-inflationary Universe
which begins on a small spatial section and which is represented
by a Penrose diagram which is somewhat taller than it is wide:
this allows all parts of a spatial section to remain in causal
contact until Inflation is ready to begin. What we have learned,
however, is that, firstly, this prescription is not unique [there
are many kinds of matter which could be born with the Universe,
and there are many possible spatial topologies], and, secondly,
the argument only works with special choices from this range of
possibilities. This in itself is not a drawback: the whole ``wave
function of the Universe" philosophy is based on the idea that the
wave function can make such choices for us. The problem lies in
trying to imagine \emph{how} this selection can be made. How can
the wave function select the appropriate matter which appears when
the Universe is born? Still more difficult: how can the wave
function express its preference for [say] the binary icosahedral
group as the fundamental group of the spatial sections? Notice
that the usual S$^4$ instanton cannot be used for this purpose
\cite{kn:gibhartle}; one would need a new instanton, possibly with
an orbifold structure\footnote{However, lest the reader gain the
impression that fundamental groups of locally spherical spaces
cannot be selected in a \emph{physical} way, let us note that in
fact there are concrete suggestions as to how it can be done: see
\cite{kn:aps}.}. Of course, one can argue that questions of this
kind will ultimately have to be confronted by \emph{any} theory of
cosmic origins; but the point is that such questions must be
answered immediately if we are to use locally spherical sections
in Linde's programme.

Although we do not regard these arguments as a conclusive
demonstration that locally spherical sections cannot be used here,
it does seem appropriate to try a simpler alternative. As was
explained in the Introduction, the size of the spatial sections is
decoupled from the spacetime curvature scale if we use flat,
compact spatial sections, so we now turn to this simpler
alternative.

\addtocounter{section}{1}
\section* {\large{\textsf{3. Global Structure of Spatially Flat Accelerating Cosmologies }}}
In this section we shall assume that our Universe was born with
\emph{flat} spatial sections, since we have seen that this is
actually a simpler procedure than using locally spherical
sections.

In fact, of course, proceeding in this way brings us into
agreement with standard practice in astrophysics. Note that some
recent observations favour spatial flatness even more strongly
than the well-known WMAP data, at the one percent level
[$\Omega_{\m{k}} = - 0.010 \pm 0.009$ \cite{kn:eisenstein}; see
however the cautionary notes in \cite{kn:sanchez}]. While such
data cannot of course rule out locally spherical spatial sections
if these are sufficiently large, and while it may be true that
some of the classical justifications for the assumption of spatial
flatness are questionable \cite{kn:lake}\cite{kn:overduin}, there
are still good theoretical reasons \cite{kn:reallyflat} to prefer
flat sections, \emph{provided} however that these are compact. The
simplest argument is that string theory abounds with extended
objects such as strings and branes: so it is natural to consider
spaces, like tori and toral quotients, in which one can place
these objects in such a way that they either cannot contract to a
point \cite{kn:brandvafa}, or can do so only with great difficulty
\cite{kn:greene}. Note in this connection that one of the more
promising approaches to string phenomenology is based on the
assumption that the ``small" dimensions correspond to spaces which
are [singular] quotients of tori \cite{kn:lust}. It is then
extremely natural to assume that the ``large" dimensions also take
the form of a [non-singular] quotient of a torus. There are in
fact many other theoretical reasons to prefer flat, compact
spatial sections, and we shall discuss some of them in the
Conclusion.

Again, it is actually standard practice in the Inflation
literature to describe de Sitter spacetime itself as having flat
spatial sections. This ``Cartesian" version of the de Sitter
metric makes use of Cartesian coordinates (t, x, y, z), running
from $-\;\infty$ to $+\;\infty$, on $\bbr^4$. The metric of
``Cartesian de Sitter" is
\begin{equation}\label{eq:H}
g(\m{CdS})(\m{L)_{+---} \;=\; dt^2\; -\;e^{(2\,t/L)}\,[\,dx^2
\;+\; dy^2 \;+\; dz^2]}.
\end{equation}
How can this metric on $\bbr^4$ be equivalent to the metric given
above on $\bbr\times\,\m{S}^3$? The short answer is that it can't:
this metric turns $\bbr^4$ into a spacetime which is timelike and
null geodesically incomplete. To see this, note that the Cartesian
coordinates (t, x, y, z) are related to the ambient Minkowski
coordinates by
\begin{eqnarray} \label{eq:F}
\m{A} & = & \m{-\;L\;sinh(t/L)\;-\;{{1}\over{2L}}\,(x^2\;+\;y^2\;+\;z^2)\,e^{t/L}  }     \nonumber \\
\m{W} & = & \m{+\;L\;cosh(t/L)\;-\;{{1}\over{2L}}\,(x^2\;+\;y^2\;+\;z^2)\,e^{t/L}}    \nonumber \\
\m{Z} & = & \m{z\,e^{t/L}}           \nonumber \\
\m{Y} & = & \m{y\,e^{t/L}}  \nonumber \\
\m{X} & = & \m{x\,e^{t/L}}.
\end{eqnarray}
Now in (\ref{eq:F}) we have
\begin{equation}\label{eq:G}
\m{W\;-\;A\;=\;L\,e^{t/L}},
\end{equation}
so W $>$ A everywhere in the domain of the coordinates. There is
no such constraint on the global conformal coordinates, and indeed
the region W $>$ A corresponds, by the first two equations in
(\ref{eq:D}), to the region $\eta\;>\;\chi$ of the Penrose
diagram, the ``upper left-hand" triangular half. That is,
``Cartesian de Sitter" is obtained by simply cutting away half of
the full de Sitter spacetime. Despite appearances, then,
\emph{this spacetime is geodesically incomplete}
\cite{kn:parikh}\cite{kn:aguirre}.
\begin{figure}[!h]
\centering
\includegraphics[width=0.8\textwidth]{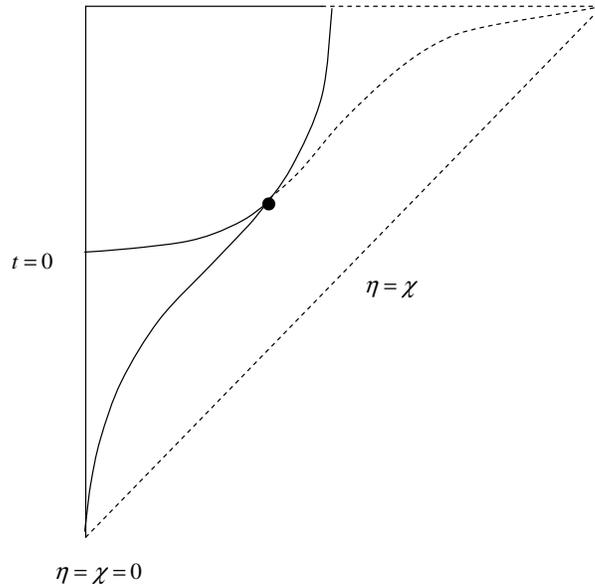}
\caption{Penrose Diagram of Toral de Sitter [solid lines].}
\end{figure}

The incompleteness of ``spatially flat de Sitter" takes on quite a
different aspect if we switch to the toral spatial topology
$\bbr\,\times\,\m{T}^3$. To see precisely how this works, take the
Cartesian coordinates and consider the hypersurface t = 0. By
equations (\ref{eq:D}) and (\ref{eq:G}), the surface t = 0
corresponds to the line in the Penrose diagram given by
\begin{equation}\label{eq:J}
\m{cos(\eta)\;+\;sin(\eta)\;=\;cos(\chi)},
\end{equation}
as shown in Figure 3.

Now take the two-sphere of radius $\pi$K in this surface; it is
represented by the heavy dot in Figure 3. This two-sphere can be
enclosed in a cube of side length 2$\pi$K. We are really
interested in this cube, but we shall discuss it in terms of the
enclosed sphere. We allow the sphere to evolve along with the
expansion: that is, we let its points trace out timelike geodesics
perpendicular to t = 0. By the last three equations in
(\ref{eq:D}) and the last three in (\ref{eq:F}), we can represent
this by the curve
\begin{equation}\label{eq:K}
\m{L\,cosec(\eta)\,sin(\chi)\;=\;\pi K\,e^{(t/L)}\;=\;\pi K\,[
\,cosec(\eta)\,cos(\chi)\;-\;cot(\eta)\,]},
\end{equation}
or
\begin{equation}\label{eq:L}
\m{cos(\eta)\;=\;cos(\chi)\;-\;{{L}\over{\pi K}}\,sin(\chi)}.
\end{equation}
This is also shown in Figure 3. If we now compactify the Cartesian
spatial section t = 0 to a cubic torus of side length 2$\pi$K, and
allow this torus to follow the geodesic shown [in the sense that
it always just encloses the two-sphere which had radius $\pi$K at
t = 0], then we have constructed spatially toral de Sitter
spacetime. Putting x = K$\theta_1$, y = K$\theta_2$, z =
K$\theta_3$ [where we recall that the angles $\theta_{1,2,3}$ run
from $-\,\pi$ to $+\,\pi$, so that the initial distance from the
centre of the cube to the boundary planes is $\pi$K], we obtain
the metric given in equation (\ref{eq:B}). To obtain the Penrose
diagram, we just have to discard all that part of Figure 3
[indicated by dotted lines] which lies to the right of the
timelike geodesic shown\footnote{Of course, this is an
approximation, valid to the extent that we are able to represent
the cube by the enclosed sphere.}. But that means that no part of
our new spacetime comes near to the line $\eta$ = $\chi$
--- \emph{except} near the single point ($\eta$ = 0, $\chi$ = 0).

Contrary to statements often found in the literature, however, the
toral version of de Sitter spacetime is, like the version with
non-compact spatial sections, \emph{geodesically incomplete}. The
only difference is that the incompleteness is now associated with
a single point ($\eta$ = 0, $\chi$ = 0). As this point is in the
infinite past according to the observers corresponding to the time
coordinate t in equation (\ref{eq:B}), it is understandable that
the incompleteness here has not been widely noticed. The fact that
toral de Sitter spacetime is incomplete nevertheless follows from
powerful singularity theorems for asymptotically de Sitter
spacetimes proved by Andersson and Galloway
\cite{kn:andergall}\cite{kn:gall}. The point is that while the
timelike geodesics associated with t are complete, \emph{other
timelike [and null] geodesics are not}. If we trace back the
worldline of a freely falling object, then we find that the total
proper time can be either infinite or finite, depending on the way
the worldline winds around the torus. Since the torus is shrinking
as we move backwards in time, the finite proper time case is easy
to achieve; in fact, that is the generic situation for timelike
geodesics, as is discussed in \cite{kn:gall}.

We conclude, then, that switching from locally spherical to flat
[compact or non-compact] spatial sections has a strong effect on
de Sitter spacetime: it forces it to become [null and timelike]
geodesically incomplete. The toral case \emph{does} represent a
great improvement over the non-compact case, but only in the sense
that objects can only enter the spacetime through a single point,
and not through the entire region $\eta$ = $\chi$. But, on the
other hand, the absence of a ``true" [curvature] singularity here
is an artifact of the very special spacetime geometry ---
\emph{generically}, the incompleteness at ($\eta$ = 0, $\chi$ = 0)
\emph{will} turn into a curvature singularity. In other words, the
toral case is more similar to a conventional cosmological model
than the non-compact case. Like a conventional cosmological model,
however, it will generically have an initial curvature
singularity.

This is of course not very welcome from the point of view of
creation from ``nothing". In fact, the situation even in the case
of compact sections is much worse than it seems, for geodesic
incompleteness is an extremely general phenomenon for spacetimes
of this kind. Indeed, one finds that the details of what is
happening in the incomplete region are almost entirely irrelevant:
the Andersson-Galloway theorems imply that, as long as a spacetime
with $\bbr\,\times\,\m{T}^3$ topology is \emph{asymptotically} de
Sitter [and non-singular at late times], then,
essentially\footnote{``Asymptotically de Sitter and non-singular
at late times" means, more precisely, that the  spacetime has a
\emph{regular future spacelike conformal boundary} and is
\emph{future asymptotically simple.}}, it \emph{has} to be
incomplete to the past as long as the \emph{Null Ricci Condition}
[see below] continues to hold. In short, we have another example
of the phenomenon which is such a remarkable feature of the
classical singularity theorems: the incompleteness is \emph{not} a
result of assumptions about symmetries, the special [FRW] form of
the metric, or even about the details of the Einstein equation.

The only way to avoid the conclusions of the Andersson-Galloway
theorems is to violate the Null Ricci Condition [NRC], which
requires that the Ricci tensor should satisfy
\begin{equation}\label{eq:M}
\mathrm{R}_{\mu\nu}\,\mathrm{k}^\mu\,\mathrm{k}^\nu\;\geq\;0.
\end{equation}
for every \emph{null} vector k$^\mu$. This condition is obviously
satisfied by de Sitter spacetime, which is why the version with
toral sections has to be incomplete. As it stands, the NRC is a
purely geometric condition, but, if we assume that the Einstein
equations hold \emph{exactly}, then it is equivalent to the
\emph{Null Energy Condition} [NEC], which is just the requirement
that the energy-momentum tensor T$^{\mu\nu}$ should satisfy
\begin{equation}\label{eq:MUG}
\mathrm{T}_{\mu\nu}\,\mathrm{k}^\mu\,\mathrm{k}^\nu\;\geq\;0
\end{equation}
for all null k$^\mu$. If, as seems at first reasonable from this
point of view, one insists that the NRC should hold, then an
asymptotically de Sitter spacetime with toral spatial sections
must be singular to the past if it is non-singular in the far
future. Typically, the Penrose diagram will either be
``triangular" --- as in Figure 3, but with the point of
incompleteness generically replaced by a genuine, curvature
singularity --- or rectangular, as in Figure 2, but with a
curvature singularity at the bottom of the diagram.

In short, generic toral cosmologies \emph{are inevitably singular
if the NRC holds.} This means that the spacetime in the
Zel'dovich-Starobinsky \cite{kn:zelda} theory \emph{will} in
reality [that is, if it contains any matter] be singular; it also,
of course, means that a toral universe cannot be created from
``nothing" if the NRC holds.

From the point of view of string theory, and specifically of
string gas cosmology \cite{kn:brandvafa}, this is a very puzzling
conclusion. For in that approach, the use of toral spatial
sections is precisely what allows the theory to \emph{avoid} an
initial singularity, through T-duality. We conclude
\cite{kn:unstable} that string gas cosmology demands a violation
of the NRC in the very earliest Universe --- that is, precisely
around the region of spacetime we are discussing here, the
pre-inflationary era --- if we follow Linde in assuming that the
spatial sections of our Universe are flat, compact
three-manifolds. All this is of course in sharp contrast to the
case of locally spherical sections. [The Andersson-Galloway
results do \emph{not} demand singularities in that case even if
the NRC holds; the technical reason being that the fundamental
groups of the sections are always \emph{finite} in that case.]

What we have learned in this section is this: if the Universe is
to be created from ``nothing" in the form of a flat three-space,
then the Universe must be born containing some structure which
violates the NRC. This structure will modify the shape of the
Penrose diagram in Figure 3 and, as we shall see, can make it
resemble Figure 2. This will of course strongly affect the
computation of the wave function of the Universe.

In order to proceed, we need some explicit examples of such
NRC-violating spacetimes. We now present a family of such
examples.

\addtocounter{section}{1}
\section* {\large{\textsf{4. NRC Violation in the Early Universe: Explicit Examples }}}
Our objective in this section is to discuss a simple family of
spacetimes describing the creation, from ``nothing", of a
spacelike torus. Later we shall also show how, and to what extent,
stringy considerations constrain its shape, but in this section
our discussion is essentially classical. We shall be guided by the
conclusions of the previous section.

The precise observational and theoretical \cite{kn:NECVIOLATION}
status of the NEC [defined above] has been much debated in recent
years. In particular, it is well known that violations of the NEC
can be dangerous: the result is often a future singularity
\cite{kn:smash}\cite{kn:kamion}. If such behaviour cannot be
avoided, then of course there can be no hope of using a spacetime
geometry with toral spatial sections as a model of the
pre-inflationary Universe. Fortunately, that is not the case, as
we shall soon see.

The first point to note is that, in contexts like the present one,
there is no reason to believe that the Einstein equation should
hold exactly, and so violations of the NRC and of the NEC
\emph{are two very different matters} here. It has been
specifically noted in the case of braneworld models
\cite{kn:varun1}\cite{kn:varun2}\cite{kn:coley} that there are
explicit corrections to the Einstein equation which allow the NRC
to be violated \emph{while every matter field satisfies the} NEC
\cite{kn:nojiri}. Similarly, the NEC and the NRC can be
significantly different in certain Gauss-Bonnet and other variants
of Einstein gravity \cite{kn:sean}\cite{kn:sasaki}. Therefore we
should not assume that violating the NRC will necessarily lead to
all of the well-known complications which may [or may not] arise
when the NEC is violated.

In order to discuss this situation concretely, we shall write the
equation governing the metric in the Einstein form, that is, as
\begin{equation}\label{eq:MONKEY}
\m{G^{\mu\nu}\;=\;8\pi
L_P^2\,[T_{Matter}^{\mu\nu}\;+\;T_{NRC}^{\mu\nu}]},
\end{equation}
where $\m{T_{NRC}^{\mu\nu}}$ represents the NRC-violating effects
we have just been discussing; in other words, if k$^{\mu}$ is
null, $\m{8\pi L_P^2\,T_{NRC}^{\mu\nu}\,k_{\mu}\,k_{\nu}}$ just
quantifies the failure of $\m{8\pi
L_P^2\,T_{Matter}^{\mu\nu}\,k_{\mu}\,k_{\nu}}$ to agree with
$\m{G^{\mu\nu}\,k_{\mu}\,k_{\nu}}$ [ =
$\m{R^{\mu\nu}\,k_{\mu}\,k_{\nu}}$]. We shall refer to the
appropriate components of this tensor, $\rho_{\m{NRC}}$ and
p$_{\m{NRC}}$, as ``energy" density and ``pressure", while bearing
in mind that these terms should not be taken literally. In
particular, these components can have unconventional signs.

In a pre-inflationary regime, one assumes that, initially, some
kind of matter is present which postpones the dominance of the
inflaton energy. There are two simple ways in which this can
happen. The first is the conventional kind of behaviour, discussed
in Section 2 above, where the density of the matter decreases with
the expansion, eventually leaving the [essentially constant]
energy of the inflaton to dominate. There is a second possibility,
however: the additional ``matter" [corresponding to
$\m{T_{NRC}^{\mu\nu}}$] could have a negative ``energy" density.
If [the absolute value of] \emph{this} density decays to zero with
the expansion, the effect will be that the total energy
\emph{increases} towards its inflationary value, instead of
decreasing towards it; but the final effect will be the same.
However, if the total energy density \emph{increases} with the
expansion, then the Hubble parameter will also increase: and this
implies that the NRC must be violated.

Our discussion of the Andersson-Galloway theorems above implies
that something of this sort \emph{must} happen if a torus is to be
created from ``nothing". This result is extremely general and does
not depend on assuming an FRW form for the metric. However, if we
do make that assumption, then we can see in detail why the NRC has
to be violated here. In this case the metric has the form
\begin{equation}\label{eq:MATRIX}
g\m{(K)_{+---} \;=\; dt^2\; -\;
K^2\;[a({{t}\over{L_{\m{inf}}}})]^2\,[d\theta_1^2 \;+\;
d\theta_2^2 \;+\; d\theta_3^2]},
\end{equation}
in an obvious generalization of equation (\ref{eq:B}); the
Friedmann equation in this case is
\begin{equation}\label{eq:MAGIC}
\m{\dot{a}^2\;=\;{{8\pi L_P^2}\over{3}}\,\rho\,a^2}.
\end{equation}
For the Universe to be created from ``nothing", there must be a
surface of vanishing extrinsic curvature. The Friedmann equation
immediately demands that the total density has to vanish
everywhere along this initial slice. Thus any positive energy
which may be present [such as that of the inflaton] must initially
be exactly cancelled by some \emph{negative} ``energy" density,
given by evaluating $\rho_{\m{NRC}}$ along the initial surface. If
$|\,\rho_{\m{NRC}}|$ now decays with the expansion, the result, as
explained above, will necessarily be a violation of the NRC.

In view of all this, we shall assume that the Universe is born,
from ``nothing", containing [a] the inflaton, [b] conventional
matter satisfying the NEC, and [c] the NRC-violating component
just discussed. Our objective now is to construct \emph{explicit}
examples spacetimes with the above properties: that is, we wish to
find spacetimes which, in violating the NRC, remain non-singular
and have a surface with the topology of a torus and with zero
extrinsic curvature. This surface should replace the point in
Figure 3 where geodesic completeness fails, and --- we hope ---
lead to a Penrose diagram like the one shown in Figure 2.

For our purposes here, it is essential to obtain an exact solution
for the spacetime metric --- the wave function of the Universe can
be computed only if we are able to evaluate the Euclidean version
of the action, and we shall need an explicit expression for the
volume form, and so on. This will of course require somewhat
drastic simplifications. The first simplification is that we shall
assume that the inflaton can be represented by a positive
cosmological constant with energy density
3/(8$\pi$L$_{\m{P}}^2$L$_{\m{inf}}^2$) and pressure
$-$3/(8$\pi$L$_{\m{P}}^2$L$_{\m{inf}}^2$). The second is that we
shall simply ignore the contribution of the conventional matter.
This is justified by the now-familiar fact that such matter only
tends to stretch the Penrose diagram vertically if it does not
cause a singularity. Including it, therefore, would only improve
our results. Finally, we shall greatly simplify our calculations
by assuming that the ``pressure" corresponding to the negative
``energy density" $\rho_{\m{NRC}}$ is given by
\begin{equation}\label{eq:MYSTERY}
\m{p_{NRC}\;=\;w_{NRC}\,\rho_{NRC}},
\end{equation}
where the ``equation-of-state parameter" $\m{w_{NRC}}$ is a
\emph{constant}, at least approximately; we believe that this
approximation, a standard one in astrophysics, does not affect any
of our conclusions.

With these three assumptions, we can now proceed in the usual way.
The vanishing of the covariant divergence of the total
stress-energy-momentum tensor leads to the conclusion that
$\rho_{\m{NRC}}$ is proportional to a$^{-\,3(1\;+\;\m{w_{NRC}})}$.
Since we want $|\,\rho_{\m{NRC}}|$ to decay with the expansion, we
must have $\m{w_{NRC}}\,>\,-\,1$. Since the inflaton energy
density 3/(8$\pi$L$_{\m{P}}^2$L$_{\m{inf}}^2$) cancels the
pressure $-$3/(8$\pi$L$_{\m{P}}^2$L$_{\m{inf}}^2$), we have
\begin{equation}\label{eq:MOUSE}
\rho\;+\;\m{p}\;=\;\rho_{\m{NRC}}\;+\;\m{p}_{\m{NRC}}\;=\;\rho_{\m{NRC}}\,[1\;+\;\m{w_{NRC}}],
\end{equation}
and then $\m{w_{NRC}}\,>\,-\,1$ implies that the NRC is violated
here, since $\rho_{\m{NRC}}$ is \emph{negative}.

It is convenient to define a constant $\gamma$ by
\begin{equation}\label{eq:MOLEHILL}
\gamma\;=\;3\,[1\;+\;\m{w_{\m{NRC}}]};
\end{equation}
from the above discussion, $\gamma$ can take any \emph{positive}
value. The presence of this new parameter is a direct result of
our need to violate the NRC: in fact, $\gamma$ measures the
maximum extent to which the NRC is violated [see below]. As we are
about to see, however, the spacetime geometry of the
pre-inflationary era is now entirely fixed, apart from this
parameter. Thus, $\gamma$ also parametrizes our ignorance of the
spacetime geometry of the earliest Universe. It will in fact turn
out that the ability of the FST wave function to fix $\gamma$ will
be crucial in realizing our scenario.

We see that $\rho_{\m{NRC}}$ is some negative multiple of
a$^{-\,\gamma}$. If we define K in equation (\ref{eq:MATRIX}) such
that the cubic torus has an initial side length 2$\pi$K, so that
the initial value of a(t) is unity, then the fact that
3/(8$\pi$L$_{\m{P}}^2$L$_{\m{inf}}^2$) must be cancelled initially
allows us to fix the constant of proportionality between
$\rho_{\m{NRC}}$ and a$^{-\,\gamma}$: evidently we must have
\begin{equation}\label{eq:MOUNTAIN}
\rho_{\m{NRC}}\;=\;{{-\,3}\over{8\pi
\m{L_P^2\,L_{\m{inf}}^2}}\,\m{a}^{\gamma}}.
\end{equation}
Adding this to the inflaton energy, substituting the total into
equation (\ref{eq:MAGIC}) and solving, we obtain a family of
metrics parametrized by $\gamma$:
\begin{equation}\label{eq:MASTODON}
g\m{(\gamma,\,K)_{+---} \;=\; dt^2\; -\;
K^2\;cosh^{(4/\gamma)}({{\gamma\,t}\over{2L_{inf}}})\,[d\theta_1^2
\;+\; d\theta_2^2 \;+\; d\theta_3^2]}.
\end{equation}
Clearly these metrics are entirely non-singular, despite the fact
that the NRC is violated everywhere: there is \emph{no} ``Big
Rip". [In fact, it was precisely for this reason that these
metrics were first introduced \cite{kn:smash}.] Furthermore, they
do all that we ask: the extrinsic curvature is zero at t = 0,
where a torus of side length 2$\pi$K is created from ``nothing";
the initial size of the torus will be small compared to the
inflationary scale as long as K/L$_{\m{inf}}$ is small. The
Universe then expands, and, since
$\m{cosh^{(4/\gamma)}({{\gamma\,t}\over{2L_{\m{inf}}}})\;\rightarrow\;e^{(2t/L_{\m{inf}})}/2^{(4/\gamma)}}$,
the metric eventually approaches the inflationary
form\footnote{See equation (\ref{eq:B}); K must be adjusted by a
factor of 2$^{2/\gamma}$.}. In fact, for the values of $\gamma$ of
interest to us here, the approach to the inflationary metric is
very rapid as measured by proper time. It proves to be much
slower, however, in \emph{conformal} time, and this is of course
vital if we are to obtain a Penrose diagram of the desired shape.

The precise geometric role of $\gamma$ may be explained as
follows: with a natural choice of null vector k$^{\mu}$ we find
that the Ricci tensor of the metric (\ref{eq:MASTODON}) satisfies
\begin{equation}\label{eq:SLEAZE}
\mathrm{R}_{\mu\nu}\,\mathrm{k}^\mu\,\mathrm{k}^\nu\;=\;\m{{{-\,\gamma}\over{
L_{inf}^2}}\,sech^2({{\gamma \, t}\over{2L_{inf}}})}.
\end{equation}
Setting t = 0, we see that $\gamma$ measures the initial [and
maximal] extent of NRC violation. For other fixed t, however, this
expression tends to zero as $\gamma$ becomes large, yet its
integral from 0 to $\infty$ is $-$2/L$_{\m{inf}}$, independent of
$\gamma$; thus the effect of taking $\gamma$ to be large is to
focus the NRC violating effect close to t = 0, while small values
of $\gamma$ describe a more ``diffuse" form of NRC violation.

One of these metrics, namely $g\m{(2,\,K)_{+---}}$, was recently
discussed \cite{kn:biswas} in connection with an attempt to
resolve the Big Bang singularity in spatially flat FRW
cosmologies, using the higher-derivative corrections of the
Einstein-Hilbert Lagrangian implied by string theory. This can be
done while avoiding ghosts, again underlining our view that
violations of the NRC need not be interpreted in terms of actual
physical fields having negative energy densities. It should be
noted that all of the technical virtues \cite{kn:biswas} of
$g\m{(2,\,K)_{+---}}$ [connected with the behaviour of
$\bigtriangleup$R, the spacetime Laplacian of the scalar
curvature] are actually shared by $g\m{(\gamma,\,K)_{+---}}$ for
\emph{all} $\gamma$. This is fortunate, since, as was discussed in
\cite{kn:unstable} [see below], $g\m{(2,\,K)_{+---}}$ itself is
not acceptable in string theory.

We can write equation (\ref{eq:MASTODON}) in the form
\begin{equation}\label{eq:MONSTROSITY}
g\m{(\gamma,\,K)_{+---}
\;=\;K^2\,cosh^{(4/\gamma)}({{\gamma\,t}\over{2L_{\m{inf}}}})\,[d\eta^2\;
-\;[d\theta_1^2 \;+\; d\theta_2^2 \;+\; d\theta_3^2]]},
\end{equation}
where $\m{cosh^{(4/\gamma)}({{\gamma\,t}\over{2L_{\m{inf}}}})}$ is
to be understood as a certain function of angular conformal time,
$\eta$. If we take one axis of the torus, say in the direction of
$\theta_1$, to be the spacelike axis, then we have a ``directed
Penrose diagram" which is a rectangle of width $\pi$ and of height
\begin{equation}\label{eq:MANGO}
\m{\Omega(\gamma,\,K)\;=\;{{2\,L_{\m{inf}}}\over{\gamma\,K}}\,\int_0^{\infty}\,{{dx}\over{cosh^{(2/\gamma)}(x)}}}.
\end{equation}
As in Figure 2, the pre-inflationary era lasts from $\eta$ = 0
until $\eta$ = $\Omega(\gamma,\,$K$)\;-\;\pi$. During this era,
the scale factor increases from unity to
$\m{cosh^{(2/\gamma)}(X)}$, where X is the solution of the
equation
\begin{equation}\label{eq:MANIAC}
\m{\pi\;=\;{{2\,L_{\m{inf}}}\over{\gamma\,K}}\,\int_X^{\infty}\,{{dx}\over{cosh^{(2/\gamma)}(x)}}}.
\end{equation}

Now our objective here, as explained in Section 2 above, is to
ensure that K is small compared to the inflationary length scale,
and that $\m{\Omega(\gamma,\,K)}$ is at least 3$\pi$ ---
preferably much larger, so that there is ample [conformal] time in
which antipodal observers can exchange signals. In fact,
$\m{\Omega(\gamma,\,K)}$ is a decreasing function of both K and
$\gamma$, and so to achieve our objectives we must understand how
\emph{both} of these parameters are fixed, not just K.

This is an essential point. It is \emph{not} true that a small
value of K/L$_{\m{inf}}$ automatically ensures that the Penrose
diagram will be tall: for that, we need also to ensure that the
terms involving $\gamma$ in (\ref{eq:MANGO}) are not too small. In
other words, the pre-inflationary spacetime geometry must also be
selected with care. Even if K indeed proves to be small,
\emph{this is of little use to us if} $\gamma$ \emph{is such that
chaotic mixing fails before the inflationary scale has been
reached.}

In fact, $\gamma$ will appear throughout our discussion, and it is
essential to understand how to constrain it. We propose that it
should be determined in much the same way as K: that is, the wave
function of the Universe should [help to] select the most probable
value of $\gamma$. The rather daunting task before us is to show
\emph{both} that K is approximately equal to the string scale
\emph{and} that $\gamma$ is such that chaotic mixing fails only
when the Universe has expanded to the inflationary scale.

The relevant Euclidean versions of our family of metrics are given
simply by complexifying time in (\ref{eq:MASTODON}), yielding
\begin{equation}\label{eq:MUMBLE}
g\m{(\gamma,\,K)_{----} \;=\; -\,dt^2\; -\;
K^2\;cos^{(4/\gamma)}({{\gamma\,t}\over{2L_{\m{inf}}}})\,[d\theta_1^2
\;+\; d\theta_2^2 \;+\; d\theta_3^2]};
\end{equation}
of course, $(-\,-\,-\,-)$ signature is equivalent to
$(+\,+\,+\,+)$. Here Euclidean time is defined on
($-\pi$L$_{\m{inf}}$/$\gamma,\; 0\,$]; at t = 0 there is the usual
transition from Euclidean to Lorentzian signature. The Euclidean
space is singular at t = $-\pi$L$_{\m{inf}}$/$\gamma$; we shall
comment on this later.

Thus we have a concrete example of a family of early-Universe
spacetime geometries which \emph{may} permit the effects Linde
requires for Inflation to begin. Furthermore, all that is required
of the wave function is that it should select two numerical
parameters, K and $\gamma$; we do not need to select the detailed
topology or to know more details of the ``matter" content. This is
of course a great simplification compared to the case of locally
spherical spatial sections.

We shall now proceed to investigate how acceptable values of K and
$\gamma$ might be derivable from the wave function of the
Universe.

\addtocounter{section}{1}
\section* {\large{\textsf{5. The Hartle-Hawking Wave Function In The Toral Case}}}
In this section we begin to explore the possibility that the basic
parameters of our model are selected by the cosmic wave function.

In the recent work [for example,
\cite{kn:tye}\cite{kn:OVV}\cite{kn:sarangi}] on creating the
Universe from ``nothing", it is assumed that the Hartle-Hawking
wave function, or a modification of it, can be used to describe
such a process. We remind the reader that this assumption has not
been accepted by all \cite{kn:lindecritical}; Linde argues that
the HH wave function describes a ground state, but \emph{not} its
creation. On the other hand, the work of Ooguri et al
\cite{kn:OVV} appears to provide evidence in favour of the
``creation interpretation", within the framework of topological
string theory. Our view is that, ultimately, a proper
interpretation [or extension] of the wave function may allow these
points of view to be reconciled. In the meantime, the most
pressing need is to see whether the ``creation interpretation" can
actually be made to work. To that end, we shall provisionally
assume that the HH and FST wave functions do predict the
probability of the creation of a Universe with specified
properties. We also assume, without further discussion, the
validity of the usual assumptions: that the ``mini-superspace"
approach is valid and that semi-classical saddle-point
approximations can be used. These standard assumptions and their
limitations are discussed in detail in
\cite{kn:carlip}\cite{kn:coule}.

According to \cite{kn:tye}\cite{kn:sarangi}, the Hartle-Hawking
wave function describes the [non-normalized] probability of
tunnelling to de Sitter spacetime, with spacetime curvature scale
L, from ``nothing". Up to constant factors this probability is
given by
\begin{equation}\label{eq:N}
\m{P_{HH}\;=\;exp(\pi \,L^2/L_P^2)},
\end{equation}
where again L$_{\m{P}}$ is the Planck length. Clearly this favours
arbitrarily large values of L. This is not acceptable physically,
and so the Hartle-Hawking wave function must be modified. The
relevant modifications will be discussed in the next section. For
the present we shall retain the Hartle-Hawking wavefunction and
explain the [rather non-trivial] procedure required to adapt it to
tunnelling to a spacetime with toral spatial sections of initial
\emph{non-zero} side length 2$\pi$K. For concreteness we shall
assume that the spacetime geometry is given by some member of the
family of metrics given by (\ref{eq:MASTODON}). Bear in mind that
the spacetime curvature length scale has a fixed value
L$_{\m{inf}}$ throughout this discussion.

We saw in the preceding Section that the relevant Euclidean
version is given by (\ref{eq:MUMBLE}): negative values of t
correspond to the Euclidean regime, while positive values of t
describe the Lorentzian geometry, and the two are to be conjoined
at t = 0. Now the Hartle-Hawking instanton with spherical spatial
sections is non-singular, even at the point where the scale factor
vanishes, because it is [part of] a four-sphere. However, this is
highly non-generic behaviour: normally one must expect the
vanishing of the scale factor to produce a singularity. In fact,
the vanishing of the Euclidean scale factor is itself no bad
thing: indeed, if the scale function did not vanish, there would
be a wormhole through to another Euclidean region, which in turn
would have further wormholes. In general this would cause the
volume of the Euclidean region, and therefore, in general, the
action, to diverge. This forces us to perform a manual cutoff at a
``plate", as in \cite{kn:sarangi}, which is not entirely
satisfactory.

In any case, the Euclidean instanton for a cosmology with toral
spatial sections \emph{has to be singular}. One can see this in
the following somewhat roundabout way. We know that the NRC has to
be violated here: this means that the sum of the total energy
density and the total pressure has to be negative. By Einstein's
equation, this means that the Hubble parameter H has to increase:
the time derivative is
\begin{equation}\label{eq:P}
\m{\dot{H}(t)\;=\;-\,4\pi\,L_P^2(\rho\;+\;p)},
\end{equation}
which is positive here. But the left side of this equation
involves two time derivatives of the scale factor. Upon
complexification, therefore, we will find that the Euclidean
version of the right side must become \emph{negative}; for
example, in the case of the metric (\ref{eq:MUMBLE}) one has
\begin{equation}\label{eq:Q}
\dot{\m{H}}(g\m{(\gamma,\,K)_{----})\;=\;-\,{{\gamma}\over{2L_{\m{inf}}^2}}\,
sec^2({{\gamma\,t}\over{2L_{\m{inf}}}})}.
\end{equation}
Therefore the Euclidean scale factor behaves exactly like the
scale factor of a FRW cosmology, with toral spatial sections,
which \emph{satisfies} the NRC. But the Andersson-Galloway results
tell us that such a cosmology has to be singular, and so the scale
factor of the original Euclidean space must vanish at some finite
t. This point \emph{is} indeed singular, because [unlike in the
case of S$^4$] the intrinsic curvature of each section is zero
here. In the case of (\ref{eq:MUMBLE}), the singularity is at t =
$-\pi$L$_{\m{inf}}$/$\gamma$.

We do not regard such a singularity as a drawback in itself; as we
have seen, the vanishing of the scale factor can help to keep the
action finite by cutting off the volume. On the other hand, a
singularity can itself cause the action to diverge. Following
Hawking and Turok \cite{kn:turokinstanton}, we shall not rule out
a singular instanton \emph{provided that it does not cause the
action to diverge}. Let us investigate this point.

First let us see what happens if we simply compute the original
Hartle-Hawking wave function for the NRC-violating geometry given
by equation (\ref{eq:MASTODON}). In the signature we are using
here, the Euclidean action of the instanton corresponding to
(\ref{eq:MUMBLE}) has the general form
\begin{equation}\label{eq:R}
\m{S_E \;=\; {{1}\over{16\pi
L_P^2}}\,\int\,R\,\sqrt{g}\,dtd\theta_1d\theta_2d\theta_3\;+\;S_{NRC}},
\end{equation}
where R is the scalar curvature and S$_{\m{NRC}}$ is the ``action"
corresponding to the NRC-violating ``matter" discussed earlier. In
a \emph{purely formal} way one can obtain an explicit expression
for this ``action" by constructing a ``phantom scalar"
\cite{kn:NECVIOLATION} model which artificially mimics the effects
of the terms involving $\rho_{\m{NRC}}$ and $\m{p_{NRC}}$ in the
Einstein equation. This is non-trivial, because we must take care
to do this in a way which is consistent with the assumption that
$\m{w_{NRC}}$ is \emph{constant}. It can however be done: if one
considers a formal field $\psi$ with a reversed kinetic term, and
uses an ``axion-like" potential, it is possible to show that the
result is an ``energy density" and ``pressure" related by a
constant which can be expressed as in equation
(\ref{eq:MOLEHILL}). [The details may be found in
\cite{kn:unstable}.] The corresponding Lagrangian proves to be a
constant multiple of
$\m{sec^2({{\gamma\,t}\over{2L_{\m{inf}}}})}$, and so by computing
the determinant of the metric given in (\ref{eq:MUMBLE}) we find
the action S$_{\m{NRC}}$:
\begin{equation}\label{eq:T}
\m{S_{NRC} \;=\;({{\gamma}\over{3}}\;-\;1)\,{{3}\over{8\pi L_P^2
L_{\m{inf}}^2}}\,\int_{-\,\pi
L_{\m{inf}}/\gamma}^0\,sec^2({{\gamma\,t}\over{2L_{\m{inf}}}})\,[8\pi^3
K^3\,cos^{(6/\gamma)}({{\gamma\,t}\over{2L_{\m{inf}}}})]\,dt}.
\end{equation}
We stress that the ``field" $\psi$ is merely a device for arriving
at this result: it plays no further role.

The scalar curvature in this signature is given by
\begin{equation}\label{eq:U}
\m{R}(g\m{(\gamma,\,K)_{----}) \;=\;
-\,{{12}\over{L_{\m{inf}}^2}}\;+\;{{3}\over{L_{\m{inf}}^2}}\,(4\;-\;\gamma)\,sec^2({{\gamma
t}\over{2L_{\m{inf}}}})},
\end{equation}
so we can compute its contribution to the action as in
(\ref{eq:T}). Combining this with (\ref{eq:T}), we find after
simplifications that the total Euclidean action becomes
\begin{equation}\label{eq:V}
\m{S_E(\gamma,\,K) \;=\; {{6\pi^2\,K^3}\over{L_{\m{inf}}\,
L_P^2\gamma}}\,\int_{-\,\pi/2}^0\,\Big\{(1\;-\;{{\gamma}\over{6}})\,
sec^2(x)\;-\;2\,\Big\}\,cos^{(6/\gamma)}(x)\,dx}.
\end{equation}
The first point to make regarding this integral is that it
diverges for all values of $\gamma\;>\;6$. \emph{This is the
result of the fact that the instanton is singular}. The integral
converges for all values less than or equal to 6, but there is
discontinuous behaviour at $\gamma$ = 6. We have in fact
\begin{eqnarray} \label{eq:W}
\m{S_E(\gamma,\,K)} & = & \m{Undefined, \;\;\gamma \;>\;6  }     \nonumber \\
\m{S_E(6,\,K)} & = & \m{{{-\,2\pi^2\,K^3}\over{L_{\m{inf}}\,L_P^2}}}    \nonumber \\
\m{S_E(\gamma,\,K)} & = &
\m{{{-\,6\pi^2\,K^3}\over{L_{\m{inf}}\,L_P^2\;\gamma}}\,\Delta_{(6/\gamma)},
\;\;\gamma\;<\;6},
\end{eqnarray}
where, for any non-negative constant $\beta$, $\Delta_{\beta}$ is
defined as
\begin{equation}\label{eq:X}
\m{\Delta_{\beta}\;=\;\int^{\pi/2}_0\,sin^{\beta}(x)\,dx}.
\end{equation}
Notice that since $\Delta_1$ = 1, $|\,\m{S_E(6,\,K)}|$ is
\emph{twice as large} as the limiting value of
$|\,\m{S_E(\gamma,\,K)}|$ as $\gamma$ approaches 6 from below.

 Leaving aside inessential factors, the Hartle-Hawking probability of
creating a Universe with toral sections and parameters $\gamma$, K
is
\begin{equation}\label{eq:Y}
\m{P_{HH}(\gamma,\,K) \;=\; e^{-\;S_E(\gamma,\,K)}}.
\end{equation}
Now one can show that $\Delta_{(6/\gamma)}/\gamma$ is a decreasing
function of $\gamma$, \emph{unbounded above} when $\gamma$ is
sufficiently small. From (\ref{eq:W}) we therefore see that the
tunnelling probability can be made arbitrarily large either by
increasing K or by decreasing $\gamma$. Various interpretations
are possible here. One is that the most probable way for the
Universe to be born is with arbitrarily large K, taking us back to
the $\bbr^3$ version of de Sitter spacetime [equation
(\ref{eq:H})]. This would leave $\gamma$ undetermined.
Unfortunately it is equally valid to interpret the result to mean
that $\gamma$ should vanish, while leaving K undetermined, which
of course does not make sense geometrically. In any case it is
clear that the wave function here is not normalisable and so its
interpretation is obscure.

Unsurprisingly --- we have not taken into account decoherence
\cite{kn:tye}\cite{kn:sarangi} effects --- the result of our
calculation is not satisfactory. Nevertheless we have learned
something important: the parameter $\gamma$ is now strongly
constrained, since the action diverges if $\gamma$ exceeds 6.
\emph{No such constraint exists classically}: $\gamma$ can take
any positive value in the metric (\ref{eq:MASTODON}). The mere
existence of the wave function imposes this constraint. The point
is that the Euclidean instanton is, as we saw, necessarily
singular in the toral case; and while this singularity has the
virtue of automatically cutting off the volume of the instanton,
the singular geometry itself will cause the action to diverge
unless $\gamma$ is constrained. The requirement that the spatial
topology should be toral, combined with the demand that the wave
function must be well-defined if the Universe is to be born at
all, strongly constrains the initial geometry. As a matter of
general principle, this is a very desirable situation: we would
hope that considerations of internal consistency should strongly
constrain or even determine the initial geometry, since this must
otherwise remain hard to explain. Thus it is a hopeful sign that
the possible range of $\gamma$ is indeed greatly reduced by such
arguments. On the other hand, the fact remains that the
Hartle-Hawking wave function seems to favour arbitrarily small
values of $\gamma$, which does not make sense. We shall have to
return to this point after considering the Firouzjahi-Sarangi-Tye
wave function in this case.

\addtocounter{section}{1}
\section* {\large{\textsf{6. The FST Wave Function In The Toral Case}}}
Sarangi and Tye \cite{kn:sarangi} stress that the impossibility of
normalizing the Hartle-Hawking wave function is in itself evidence
that the wave function has to be modified. Ideally, the requisite
modification here should be derived from string theory. String
theory does have some very general properties which are surely
relevant, and we can therefore attempt to guess the general form
the modifications must take. For example, in string theory
spacetime is ultimately 10-dimensional, and there is a natural
length scale, the string scale L$_{\m{s}}$. These general facts
give us some guidance as to the form of a modified wave function.
On the other hand, it is stressed by Firouzjahi, Sarangi, and Tye
\cite{kn:tye}\cite{kn:sarangi} that the wave function must also be
shaped by quantum decoherence induced by an ``environment"
consisting of perturbative modes of the spacetime geometry,
together with whatever matter is created in the beginning.

Combining these basic observations, we expect a correction to the
Hartle-Hawking wave function which depends on the 10-dimensional
volume of the Euclidean instanton, measured in string units.
Firouzjahi et al begin with a spacetime with locally spherical
spatial sections and spacetime length scale L; the specific form
they propose for the non-normalized probability is
\begin{equation}\label{eq:BETA}
\m{P_{FST}\;=\;exp\Big({{\pi
\,L^2}\over{L_P^2}}\;-\;c\,{{V_{10}}\over{L_s^{10}}}\Big)},
\end{equation}
where L$_{\m{s}}$ is the string length scale, V$_{10}$ is the
ten-dimensional volume corresponding to the Euclidean instanton
describing the tunnelling of the Universe from ``nothing", and c
is a constant which is to be calculated [from quantum decoherence
theory] given the precise details of the vacuum. Since V$_{10}$
depends on L, the presence of two competing factors in the
expression for $\m{P_{FST}}$ immediately means that arbitrarily
large values of L may no longer be preferred, and indeed
Firouzjahi et al are able to argue, beginning with de Sitter
spacetime with spherical sections, that the ``KKLMMT" inflationary
scenario \cite{kn:KKLMMT} is the most probable way for the
Universe to be born. Our objective here is to see what the
modified wave function predicts if we assume that the Universe is
born, instead, with \emph{toral} spatial sections. For
concreteness we shall assume that the spacetime geometry is
described by some member of the family of metrics given by
(\ref{eq:MASTODON}), leaving the parameter $\gamma$ undetermined
for the moment. As usual, we fix L at L$_{\m{inf}}$ henceforth,
since the initial size of the Universe is determined by K, not L.

The four-dimensional volume of the instanton is easily computed:
\begin{equation}\label{eq:GAMMA}
\m{V_4\;=\;8\,\pi^3\,K^3\,\int_{-\,\pi\,L_{\m{inf}}
/\gamma}^0\,cos^{(6/\gamma)}({{\gamma\,t}\over{2L_{\m{inf}}}})\,dt\;=\;{{16\,\pi^3\,K^3\,L_{\m{inf}}}
\over{\gamma}}\,\Delta_{(6/\gamma)}.}
\end{equation}
It is very natural to assume here that the internal space is a
toral orbifold, as for example in \cite{kn:lust}. In our
three-torus, 2$\pi$K can be defined as the length of the shortest
closed geodesic: this is the most natural definition, in view of
the importance of \emph{circumnavigations} in our whole approach.
Now clearly we have to find an explicit formulation of the
intuitive idea that the initial size of all dimensions should be
the ``same", as in string gas cosmology \cite{kn:brandvafa}. In
the case of the orbifold, it is not possible to insist on this
literally, but we can argue that the \emph{average} size of the
internal dimensions should be the same as the initial size of the
three dimensions which are destined to become large: that is, it
should be about 2$\pi$K. As this is the length of the shortest
closed geodesic on the three-torus, and since --- again as in
string gas cosmology --- T-duality leads us to expect a minimum
effective size for all dimensions, one might argue that we should
say that the size of the internal dimensions should be \emph{at
least} 2$\pi$K, but let us be conservative and use this value.
Then the internal volume is well approximated by [2$\pi$K]$^6$.
Thus we have
\begin{equation}\label{eq:DELTA}
\m{V_{10}\;=\;{{[2\,\pi]^9\,\,K^9\,L_{\m{inf}}}\over{\gamma}}\,\Delta_{(6/\gamma)},}
\end{equation}
and so the FST wave function in our case yields a probability
function
\begin{equation}\label{eq:EPSILON}
\m{P_{FST}(\gamma,\,K)\;=\;exp\Big\{-\;S_E(\gamma,\,K)\;-\;c\,{{[2\,\pi]^9\,\,K^9\,L_{\m{inf}}}\over
{\gamma\,L_s^{10}}}\,\Delta_{(6/\gamma)}\Big\}}.
\end{equation}
According to the equations (\ref{eq:W}) this is
\begin{eqnarray} \label{eq:ZETA}
\m{P_{FST}(\gamma,\,K)} & = & \m{Undefined, \;\;\gamma \;>\;6  }     \nonumber \\
\m{P_{FST}(6,\,K)} & = &
\m{exp\Big\{\,{{2\pi^2\,K^3}\over{L_{\m{inf}}\,L_P^2}}\;-\;c\,{{[2\,\pi]^9\,\,K^9\,L_{\m{inf}}}\over
{6\,L_s^{10}}}\Big\}}    \nonumber \\
\m{P_{FST}(\gamma,\,K)} & = &
\m{exp\Big\{\,\Big[\,{{6\pi^2\,K^3}\over{L_{\m{inf}}\,L_P^2}}\;-\;c\,{{[2\,\pi]^9\,\,K^9\,L_{\m{inf}}}\over
{L_s^{10}}}\,\Big]{{\Delta_{(6/\gamma)}}\over{\gamma}} \Big\}}
,\;\;\gamma\;<\;6.
\end{eqnarray}

We see at once that large values of K are no longer favoured, and
this is of course highly desirable. Equally, however, we see that
the FST wave function does not in itself resolve the $\gamma$
problem: for any given value of K, we can make
$\m{P_{FST}(\gamma,\,K)}$ arbitrarily large simply by choosing
$\gamma$ to be sufficiently small. Again, this does not make sense
either geometrically [equation (\ref{eq:MASTODON})] or in terms of
having a normalizable wave function.

It is worth stressing this point. One might have thought that
generalizing the FST wave function to the toral case would be a
straightforward matter: the wave function should favour a
particular value of K here, just as it favours a particular value
of L in the locally spherical case. This proves to be incorrect.
The ultimate reason for this is again the Andersson-Galloway
results, which tell us that the toral topology will lead to a
singularity unless we violate the NRC. But violating the NRC
forces us to introduce a new parameter, $\gamma$, which itself
strongly affects the structure of the wave function. One can
certainly expect difficulties like this to arise in general, not
just for the particularly simple family of metrics we have used
here.

As we have emphasised, classically there is no constraint on
$\gamma$ other than that it should be positive: any value makes
sense in equation (\ref{eq:MASTODON}). But we saw that internal
consistency --- the requirement that the wave function should be
well-defined --- imposes a strong constraint on $\gamma$ when the
quantum theory is constructed. This constraint is the \emph{upper}
bound $\gamma\;\leq\;6$. Our only hope now is that similar
self-consistency arguments can impose a \emph{lower} bound on
$\gamma$. This is in fact precisely what happens, as we now
explain.

\addtocounter{section}{1}
\section* {\large{\textsf{7. How Non-Perturbative String Physics Constrains $\gamma$}}}

We argued earlier that it is necessary for the NRC to be violated
if the Universe is to be created from ``nothing" as a torus.
However, we also argued that this could be done without violating
the Null Energy Condition [as opposed to the Null \emph{Ricci}
Condition]. It might seem, therefore, that violating the NRC may
not have any physical consequences provided that we decouple the
NRC from the NEC.

In \cite{kn:unstable} it was pointed out, however, that in string
theory there is a known source of \emph{non-perturbative}
instability \emph{which is related only to the} NRC, \emph{not to
the} NEC. This is the specifically stringy process discovered by
Seiberg and Witten \cite{kn:seiberg}. Take a BPS (D $-$ 1)-brane
together with an appropriate antisymmetric tensor field in a
manifold which is asymptotic to the (D + 1)-dimensional hyperbolic
space H$^{\mathrm{D} + 1}$. The brane action takes the form
\cite{kn:seiberg}
\begin{equation}\label{eq:Z}
\mathrm{S_B} \;=\;
\mathrm{T}(\mathrm{A}\;-\;{{\mathrm{D}}\over{\mathrm{L}}}\,\mathrm{V}),
\end{equation}
where T is the tension of the brane, A is its area, V the volume
enclosed, and L is the length scale of the asymptotic hyperbolic
space. The point is that this action is a purely geometric object.
If the geometry of the ambient space is such that [for example]
this action is unbounded below, then the result will be a severe
instability due to the nucleation of ``large branes", \emph{no
matter how} the geometry came to have that particular shape.

In fact, Seiberg and Witten were able to show that this
non-perturbative\footnote{The instability is non-perturbative in
the sense that a barrier has to be overcome to create the branes.}
stringy instability is never a problem if the conformal structure
at infinity is represented by a metric of positive scalar
curvature; on the other hand, it is unavoidable if the scalar
curvature at infinity is negative. The case of scalar-flat
[including, of course, completely flat] boundaries is particularly
delicate: Seiberg-Witten instability occurs in some cases but not
in others\footnote{A particularly clear discussion of this and of
related issues is given by Kleban et al in \cite{kn:porrati}.}.

In \cite{kn:unstable}, it was found that NRC-violating spacetimes
with flat compact spatial sections frequently [not invariably] do
suffer from Seiberg-Witten instability: that is, string theory
frequently rules out models which appear to be classically
consistent. In particular, if we consider the asymptotically
hyperbolic version of any spacetime of the form given in equation
(\ref{eq:MATRIX}), we find that \emph{the brane action is always
unbounded below whenever the NRC is violated and the NRC-violating
term decays towards infinity at the same rate as, or more slowly
than,} a(t)$^{-\,3}$. From equations (\ref{eq:MOUSE}) and
(\ref{eq:MOUNTAIN}) we see that this just means that values of
$\gamma$ less than or equal to 3 are absolutely forbidden here.
This is precisely the lower bound we need.

The situation for values of $\gamma$ greater than 3 but less than
6 is much more delicate. The action for a Seiberg-Witten ``large
brane" in the relevant geometry here,
$\m{S_B}(\gamma,\,\m{K,\,t})$, is always positive at t = 0, but it
then decreases; it is bounded below if $\gamma\;>\;3$, but the
limiting value is given by \cite{kn:unstable}
\begin{equation}\label{eq:ZORILLO}
\lim_{\m{t} \rightarrow \infty}{\m{S_B}(\gamma,\,\m{K,\,t)}}\;=\;
2^{(3\;-\;{{6}\over{\gamma}})}\,\pi^3\,\m{K^3\,T}\,
{{(\gamma\;-\;6)(3\;+\;\gamma)}\over{\gamma(\gamma\;-\;3)}}.
\end{equation}
We immediately see that the \emph{largest} value permitted by the
existence of the wave function, 6, is the \emph{smallest} value
which guarantees that the brane action should never be negative.
This is very remarkable, and certainly suggests that $\gamma$ = 6
is the preferred value. For values strictly between 3 and 6, the
brane action does become negative for sufficiently large t, and it
continues to decrease thereafter: there is no minimum for any
finite t. This means that we can create a brane/antibrane pair
and, by moving them to sufficiently large t, we can reduce the
action. Admittedly the reduction is only by a finite amount, but,
as in the similar cases discussed by Maldacena and Maoz in
\cite{kn:maoz}, there is a danger of non-perturbative instability
for these values of $\gamma$ also.

This suggests that we should exclude \emph{all} values of $\gamma$
below 6, leaving 6 as the only physically acceptable value of
$\gamma$. However, that may be premature. Notice that the brane
action becomes negative only beyond a certain critical value of t,
t$_{\m{c}}$, which depends on $\gamma$. If $\gamma$ is close to 6,
then t$_{\m{c}}$ can be very large, and indeed it may occur beyond
the end of the pre-inflationary period. But in that region we have
no reason to believe that the NRC continues to be violated;
indeed, since our assumption is that the NEC is satisfied at all
times, and since the Einstein equations are presumably
approximately valid during the inflationary era, we must expect
that NRC violation ceases at some point. At that point, the brane
action may begin to increase, so it may never have become
negative. Thus we should be cautious about excluding values of
$\gamma$ near to [but less than] 6.

A detailed investigation shows that this argument still excludes
most values of $\gamma$ up to a value quite close to 6. Rather
than pursue these details, however, we shall find it more
instructive to allow $\gamma$ to vary in the range
$3\;<\;\gamma\;\leq\;6$. The wave function itself will select the
appropriate value of $\gamma$ in this range.

We can summarize the findings of this section very briefly.
Classically, the crucial parameter $\gamma$ can range between 0
and infinity. The self-consistency of the ``creation from nothing"
scenario requires $\gamma$ to be no greater than 6; the
self-consistency of a string-theoretic formulation of the relevant
spacetime geometry requires $\gamma$ to be greater than some value
below, but quite close to, 6. With these constraints we can fix
the most probable value of $\gamma$ and therefore of K.

\addtocounter{section}{1}
\section* {\large{\textsf{8. The Most Probable Values of K and $\gamma$}}}

Let us finally compute the most probable values of our two basic
parameters, using both the Hartle-Hawking and the FST wave
functions.

According to the HH wave function, $|\,\m{S_E(\gamma,\,K)}|$
measures the probability of tunnelling with given values of K and
$\gamma$. By the equations (\ref{eq:W}) we know that the HH wave
function favours arbitrarily large values of K. Let us assume a
cutoff at some value K$_{\m{cut}}$ of K which is very large
compared to L$_{\m{P}}$ and L$_{\m{inf}}$ [and independent of
$\gamma$], and ask which value of $\gamma$ is preferred within the
range $(3,\;6]$ discussed earlier. Now
$\Delta_{(6/\gamma)}/\gamma$ decreases everywhere on $(0,\;6)$,
but recall the curious fact that the action is
\emph{discontinuous} at 6. To clarify the consequences of this, we
compare three values or limiting values of
$|\,\m{S_E(\gamma,\,K_{\m{cut}})}|$ to see where the maximum lies.
We find that
\begin{eqnarray} \label{eq:KAPPA}
\m{{{L_{\m{inf}}\,L_P^2}\over{2\pi^2\,K_{\m{cut}}^3}}}\;\m{|\,S_E(6,\,K_{\m{cut}})|}
& = & 1
\nonumber \\
\m{{{L_{\m{inf}}\,L_P^2}\over{2\pi^2\,K_{\m{cut}}^3}}}\;\m{|\,S_E(3,\,K_{\m{cut}})|}
& = &
\pi/4 \nonumber\\
\m{{{L_{\m{inf}}\,L_P^2}\over{2\pi^2\,K_{\m{cut}}^3}}}\;\lim_{\gamma\,\rightarrow\,6}\;\m{|\,S_E(\gamma,\,K_{\m{cut}})|}
& = & 1/2.
\end{eqnarray}
Here, as agreed in the previous Section, we have allowed $\gamma$
to go as low as 3. Evidently $\gamma$ = 6 is the favoured value.
The \emph{extent} to which it is preferred is measured by the
fractions
\begin{eqnarray} \label{eq:LAMBDA}
\m{P_{HH}(6,\,K_{\m{cut}})/P_{HH}(3,\,K_{\m{cut}})} & = & \m{
exp\,\Big\{{{\,2\pi^2\,K_{\m{cut}}^3}\over{L_{\m{inf}}\,L_P^2}}\,[1\;-\;{{\pi}\over{4}}]}\Big\} \nonumber \\
\m{P_{HH}(6,\,K_{\m{cut}})/\lim_{\gamma\,\rightarrow\,6}\,P_{HH}(\gamma,\,K_{\m{cut}})
} & = & \m{
exp\,\Big\{{{\,\pi^2\,K_{\m{cut}}^3}\over{L_{\m{inf}}\,L_P^2}}\Big\}}.
\end{eqnarray}
Since we are assuming that K$_{\m{cut}}$ is much larger than
L$_{\m{P}}$ and L$_{\m{inf}}$, both numbers on the right here are
enormous. Thus the Hartle-Hawking wave function strongly favours
$\gamma$ = 6, reconfirming the [of course, entirely independent]
argument in the previous section, which was based on the
self-consistency of the string theory formulation.

Although this calculation was unacceptably vague, in that the HH
wave function does not determine K, it does illustrate two points.
The first is related to the simple fact that creating the Universe
is obviously not an experiment that we can perform more than once.
Hence it is vital that all probability distributions in such
discussions should be very strongly peaked; otherwise we should
not know how to interpret the results. The second point is that
\emph{discontinuities in the probability function can be useful},
for they may allow us to make very precise predictions. In the
present case, for example, $\gamma$ = 6 is enormously more
probable than $\gamma$ = 5.99999. One should look for other
examples of this sort.

Now let us turn to the FST wave function. Here the discontinuity
complicates the analysis, since the favoured value of K is
different for $\gamma$ = 6 and values of $\gamma$ arbitrarily
close to 6. From the equations (\ref{eq:ZETA}) we see that K$^*$,
the most probable value of K, is either the value K$^*_6$  that
maximizes $\m{P_{FST}(6,\,K)}$ or the value K$^*_3$ that maximizes
$\m{P_{FST}(3,\,K)}$. [Notice that this same value, K$^*_3$,
actually maximizes $\m{P_{FST}(\gamma,\,K)}$ for each fixed
$\gamma$ strictly less than 6.] One finds
\begin{equation}\label{eq:THETA}
\m{\Big({{K_6^*}\over{L_{\m{inf}}}}\Big)^6\;=\;2\,\Big({{K_3^*}\over{L_{\m{inf}}}}\Big)^6\;=\;
{{1}\over{(2\pi)^7}}\,{{1}\over{c\,}}\,
\Big({{L_s}\over{L_P}}\Big)^2\,\Big({{L_s}\over{L_{\m{inf}}}}\Big)^8.
}
\end{equation}
Substituting (\ref{eq:THETA}) back into (\ref{eq:ZETA}) we can
compute ratios analogous to those in (\ref{eq:LAMBDA}) above:
\begin{eqnarray} \label{eq:MU}
\m{{{P_{FST}(6,\,K_6^*)}\over{P_{FST}(3,\,K_3^*)}}} & = & \m{
exp\,\Big\{{{1}\over{3\,\sqrt{c}\,(2\pi)^{3/2}}}\,[1\;-\;{{\pi}\over{4\sqrt{2}}}]
\Big({{L_s}\over{L_P}}\Big)^3\,\Big({{L_s}\over{L_{\m{inf}}}}\Big)^2}\Big\} \nonumber \\
\m{{{P_{FST}(6,\,K_6^*)}\over{\lim_{\gamma\,\rightarrow\,6}\,P_{FST}(\gamma,\,K_3^*)}}
} & = & \m{ \m{
exp\,\Big\{{{1}\over{3\,\sqrt{c}\,(2\pi)^{3/2}}}\,[1\;-\;{{1}\over{2\sqrt{2}}}]
\Big({{L_s}\over{L_P}}\Big)^3\,\Big({{L_s}\over{L_{\m{inf}}}}\Big)^2}\Big\}}.
\end{eqnarray}

Obviously, several quantities in these relations are not known
exactly; fortunately, however, the results are not very sensitive
to the details of our choices. Reasonable estimates are as
follows. First, Firouzjahi et al \cite{kn:tye} estimate c at about
10$^{-\,3}$; although the value is somewhat model-dependent, we
shall follow them and use this value. [We shall see that our
prediction for the initial size differs considerably from that of
FST; by choosing the same value of c, we can make it clear that
this is not due to a different choice of parameters.] Next, we
shall assume that the string scale is about two orders of
magnitude below the Planck scale and that the inflationary scale
is about two orders of magnitude lower again. Before proceeding,
we emphasise that these values, or any values not drastically
different, cause the functions $\m{P_{FST}(6,\,K)}$ and
$\m{P_{FST}(3,\,K)}$ to become extremely sharply peaked about
their maxima: even a 10\% variation in K away from the preferred
value causes them to decrease by many orders of magnitude. Thus,
once again, although the predictions made by the wave function are
probabilistic, they are very robust. This is clear also in the
case of $\gamma$: with these data, we find
\begin{eqnarray} \label{eq:NU}
\m{{{P_{FST}(6,\,K_6^*)}\over{P_{FST}(3,\,K_3^*)}}} & \approx &
8.4\;\times\;10^{12} \nonumber \\
\m{{{P_{FST}(6,\,K_6^*)}\over{\lim_{\gamma\,\rightarrow\,6}\,P_{FST}(\gamma,\,K_3^*)}}
} & \approx & 6.2\;\times\;10^{18}.
\end{eqnarray}
As before, $\gamma$ = 6 is favoured over all other admissible
values of $\gamma$, including [especially] values close to 6; and
evidently the prediction $\gamma$ = 6 is sufficiently robust. We
can now use $\gamma$ = 6 in (\ref{eq:THETA}) to compute K$^*$ =
K$^*_6$.

We find that K$^*$/L$_{\m{inf}}$ is about 1/270. Substituting this
into the formula (\ref{eq:MANGO}) for the height of the Penrose
diagram, using $\gamma$ = 6, we find
\begin{equation}\label{eq:IOTA}
\m{\Omega(6,\,K^*)\;\approx\;330}.
\end{equation}
Recalling that the width of the diagram is $\pi$, we see that the
most probable shape for the Penrose diagram in the
pre-inflationary phase is a rectangle about 100 times as high as
it is wide. This is ample to allow for chaotic mixing during that
phase.

Solving equation (\ref{eq:MANIAC}) in this case, we have X
$\approx$ 14.05. Thus during the pre-inflationary era the torus
grows from an initial side length 2$\pi$K$^*\,\approx$
0.023L$_{\m{inf}}$ --- \emph{that is, roughly the string scale}
--- to 2$\pi$K$^*\,$cosh$^{(1/3)}$(14.05) $\approx$
2L$_{\m{inf}}$; that is, Inflation begins at about the time global
causal contact is lost, as we hoped.

We have given evidence that the FST wave function does naturally
predict Linde's \cite{kn:lindetypical} scenario. In detail,
assuming that the Universe tunnels from ``nothing" in the form of
a torus
--- other flat three-manifolds with more complicated topologies
would give similar results --- we find that the \emph{most
probable} initial side length is about the size of the string
scale, in agreement with string gas cosmology \cite{kn:brandvafa}.
The torus then expands through a pre-inflationary phase, during
which all parts of the Universe remain in causal contact. The
spacetime geometry during this phase, granted the many
simplifications we have assumed, is \emph{most probably} described
by a metric of the approximate form
\begin{equation}\label{eq:OMICRON}
g\m{(6,\,{{L_{inf}}\over{270}})_{+---} \;=\; dt^2\; -\;
\Big({{L_{inf}}\over{270}}\Big)^2\;cosh^{(2/3)}({{3\,t}\over{L_{inf}}})\,[d\theta_1^2
\;+\; d\theta_2^2 \;+\; d\theta_3^2]}.
\end{equation}
This phase ends roughly when global causal contact is lost, by
which time the torus is of about the size of the usual
inflationary length scale. At that time, the appropriate
conditions for Inflation to begin are satisfied, and then the
evolution proceeds as usual.

 \addtocounter{section}{1}
\section*{\large{\textsf{9. Conclusion }}}

Motivated by string gas cosmology \cite{kn:brandvafa}, we have
proposed a model in which the Universe is created, from
``nothing", in the form of a torus. This, through the FST wave
function, allows a self-consistent account of the beginning of
Inflation.

Naturally we do not claim that the model presented here is
realistic, and indeed much would need to be done to make it so.
One would need to work with a more precise version of the FST wave
function, to include the effects of conventional matter, to
account for the details of the end of the NRC-violating effects
[``crossing the Phantom divide", in Hu's \cite{kn:hu} terminology;
see \cite{kn:unstable}\cite{kn:crossing}], to give a more detailed
description of the NRC-violating effects and use it to derive a
more realistic metric, and so on. Nevertheless there are many
aspects of our discussion which can be expected to survive in a
more realistic formulation. For example, the inclusion of
conventional matter can be expected, in accordance with the
Gao-Wald theorem \cite{kn:gaowald}, to help to make the
pre-inflationary Penrose diagram somewhat \emph{taller}; so we can
expect that our prediction --- of pre-inflationary Penrose
diagrams that are much taller than they are wide --- is robust.
Again, the Andersson-Galloway results are independent of our
special simplifying assumptions, so it is clear that NRC violation
in the very early Universe is generic if we assume that the
Universe tunnelled from ``nothing" to a torus. But NRC violation
will be described by some generalized version of the parameter
$\gamma$ discussed in this paper, and these new parameters will
have to be fixed in some way. Our discussion here suggests that
this will be accomplished by means of a combination of string
dynamics and probabilistic arguments based on the wave function
itself. These are the general, and most important, lessons of this
simplified investigation.

We have noted that controversy continues as to the real status of
the Hartle-Hawking and FST wave functions \cite{kn:lindecritical}.
In particular, Linde objects that these wave functions should not
be interpreted, as for example by Firouzjahi et al \cite{kn:tye}
and Ooguri et al \cite{kn:OVV}, in terms of ``probabilities of
creation" at all; indeed, in \cite{kn:lindetypical} he puts
forward the intriguing suggestion that they have to do with
\emph{final} rather than with initial conditions. Linde's basic
objection is that, physically, it is not reasonable that it should
be easier to create a larger universe than a smaller one. Now the
HH wave function suggests that the Universe should be born
arbitrarily large; the FST wave function, applied to cosmologies
with locally spherical sections, suggests that it should be born
at about the inflationary scale; and we have argued that, applied
to cosmologies with compact flat sections, the preferred length
scale is still smaller, about the string scale. In view of this
tendency towards smaller scales, it is natural to ask whether some
further refinement might indeed bring the predicted length scale
down to the Planck scale, as Linde prefers. This may be possible
if the FST parameter c, which is determined by details of
decoherence physics \cite{kn:sarangi}, can be made very much
larger [see equation (\ref{eq:THETA})] than the value assumed here
and in \cite{kn:tye}; one may also need to assume a smaller value
for the string length scale. These may not seem very plausible
assumptions, but one should bear in mind that the ``matter" being
considered here is very unusual. Clearly this point merits further
investigation.

Our combination of Linde's proposal with that of FST pushes the
tunnelling event to [or maybe even below] the string length scale,
and so it brings to the fore, once again, the difficult question
of how cosmology can be reconciled with the basic machinery of
string theory: in particular, with the existence of an S-matrix.
Following Bousso \cite{kn:bousso}, we wish to stress that, in
cosmological applications, non-compact spatial sections are not
particularly helpful if we wish to take an S-matrix point of view.
In fact, they can make the puzzle worse. Consider for example the
decelerating FRW models with \emph{non-compact} flat spatial
sections. These have a spacelike Big Bang singularity, and an
observer at any finite time can be aware of only an ``infinitely
negligible" fraction of the information stored on a spatial slice.
In this situation, apart from any other difficulties, one has to
deal with the fact that the initial state is almost entirely
unknown to any given observer. By contrast, if the spatial
sections are tori, \emph{and if the geometry is such that an
entire section is visible at a finite time} during the
pre-inflationary era, then a given ``observer" can be fully
informed of the state of the entire Universe by the time Inflation
starts.

It seems then that the picture of a globally causally connected
pre-inflationary era may throw some light on Bousso's concerns.
But this will only work if K is sufficiently small and if the
pre-inflationary geometry has the right form. Thus we have another
strong motive to investigate the possibility that the FST wave
function might naturally favour values of K which are small
relative to the spacetime curvature scale, and spacetime
geometries which allow global causal contact.

Other reasons for thinking that compact flat spatial sections may
be directly relevant to string cosmology are given in
\cite{kn:silver}: the existence or non-existence of winding modes
for tachyons means that the physics in the case of toral spatial
sections is very different from the spherical case. This may in
turn prove relevant to our concerns, since \cite{kn:silver}
indicates that there may be a specifically stringy version of
creating the Universe from ``nothing".

We have avoided any discussion of the difficult questions
surrounding the state of the Universe in its earliest moments
\cite{kn:sorbo}\cite{kn:chen}. We believe that these questions can
only be answered by string theory, perhaps in the guise of string
gas cosmology \cite{kn:brandlook}. It is natural to hope that the
FST ideas will be helpful here, and it is a good sign that they
are compatible with the string gas approach.

\addtocounter{section}{1}
\section*{\textsf{Acknowledgements}}
The author is deeply grateful to Wanmei for preparing the diagrams
--- despite being extremely busy --- and for being herself. He also
sincerely thanks all those who have made possible his visit to the
High Energy, Cosmology, and Astroparticle Physics Section of the
Abdus Salam International Centre for Theoretical Physics, where
the work described here was initiated. He has also benefited from
very helpful correspondence with Profs Galloway and Linde.

\end{document}